\documentclass[10pt,journal]{IEEEtran}
\def\BibTeX{{\rm B\kern-.05em{\sc i\kern-.025em b}\kern-.08em
    T\kern-.1667em\lower.7ex\hbox{E}\kern-.125emX}}
\usepackage{amsmath}
\usepackage{amsthm}
\usepackage{amsfonts}
\usepackage{amssymb}
\usepackage{graphicx}
\usepackage{cite}
\usepackage{subfigure}
\usepackage{balance}
\usepackage{algorithm}
\usepackage{algorithmic}
\usepackage{enumerate}
\usepackage{color}
\usepackage{array}
\usepackage{indentfirst}
\usepackage{multirow}
\usepackage{booktabs}
\usepackage{color}
\usepackage{slashbox}
\usepackage{diagbox}
\usepackage{threeparttable}
\usepackage[svgnames,table]{xcolor}
\usepackage{bm}
\usepackage{bbm}
\usepackage{subfigure}
\usepackage{epstopdf}

\makeatletter
\newcommand{\tb}{\mathbf}

\newcommand{\Rmnum}[1]{\expandafter\@slowromancap\romannumeral #1@}
\makeatother

\begin{document}
 	\title{{Learning-based Predictive Beamforming for Integrated Sensing and Communication \\ in Vehicular Networks}}


\author{\IEEEauthorblockN{Chang Liu, \emph{Member, IEEE}, Weijie Yuan, \emph{Member, IEEE}, Shuangyang Li \emph{Member, IEEE}, \\ Xuemeng Liu, \emph{Graduate Student Member, IEEE}, Husheng Li, \emph{Senior Member, IEEE}, \\ Derrick Wing Kwan Ng, \emph{Fellow, IEEE}, and Yonghui Li, \emph{Fellow, IEEE} }


\thanks{C. Liu, S. Li, and D. W. K. Ng are with the School of Electrical
Engineering and Telecommunications, University of New South Wales, Sydney, Australia a (e-mail: chang.liu19, shuangyang.li, w.k.ng@unsw.edu.au).}

\thanks{W. Yuan is with the Department of Electrical and Electronic Engineering, Southern University of Science and Technology, Shenzhen, 518055, China (e-mail: yuanwj@sustech.edu.cn).}

\thanks{X. Liu and Y. Li are with the School of Electrical and Information Engineering, University of Sydney, Sydney, NSW 2006, Australia (e-mail: xuemeng.liu, yonghui.li@sydney.edu.au).}

\thanks{H. Li is with the Department of Electrical Engineering and Computer Science, University of Tennessee, Knoxville, TN 37996 USA (e-mail: hli31@utk.edu).}

\thanks{This work has been accepted in part for presentation at the IEEE International Conference on Communications (ICC), 2022.}


}

\maketitle

\begin{abstract}
This paper investigates the integrated sensing and communication (ISAC) in vehicle-to-infrastructure (V2I) networks.
To realize ISAC, an effective beamforming design is essential which however, highly depends on the availability of accurate channel tracking requiring large training overhead and computational complexity.
Motivated by this, we adopt a deep learning (DL) approach to implicitly learn the features of historical channels and directly predict the beamforming matrix to be adopted for the next time slot to maximize the average achievable sum-rate of an ISAC system.
The proposed method can bypass the need of explicit channel tracking process and reduce the signaling overhead significantly.
To this end, a general sum-rate maximization problem with Cramer-Rao lower bounds-based sensing constraints is first formulated for the considered ISAC system taking into account the multiple access interference.
Then, by exploiting the penalty method, a versatile unsupervised DL-based predictive beamforming design framework is developed to address the formulated design problem. As a realization of the developed framework, a historical channels-based convolutional long short-term memory (LSTM) network (HCL-Net) is devised for predictive beamforming in the ISAC-based V2I network. Specifically, the convolution and LSTM modules are successively adopted in the proposed HCL-Net to exploit the spatial and temporal dependencies of communication channels to further improve the learning performance.
Finally, simulation results show that the proposed predictive method not only guarantees the required sensing performance, but also achieves a satisfactory sum-rate that can approach the upper bound obtained by the genie-aided scheme with the perfect instantaneous channel state information available.
\end{abstract}


\section{Introduction}
The next generation wireless systems are expected to provide not only ultra high-speed communication rate, but also high-accuracy sensing services \cite{liu2020joint}. Conventionally, communication and sensing systems are allocated to different orthogonal frequency bands and designed independently. Recently, the rapid development of multi-antenna technologies, especially the massive multiple-input multiple-output (MIMO) and millimeter wave (mmWave) technologies, grants future communication systems the capability to also perform high-accuracy sensing tasks. As such, the notion of integrated sensing and communication (ISAC) \cite{liu2021survey, akan2020internet}, in which sensing and communication systems are co-designed to share the same frequency band and hardware, has been proposed as an enabling technology for beyond fifth-generation (5G) and sixth-generation (6G) wireless systems to further improve the spectral efficiency and to reduce the hardware-cost \cite{wong2017key}.
It is envisioned that the ISAC can support various essential applications \cite{liu2021survey}, ranging from indoor localization, extended reality, to unmanned aerial vehicle (UAV) sensing and communication.
Among various emerging network architectures, vehicle-to-everything (V2X) networks \cite{wymeersch20175g} play an important role to unlock the potential of upcoming next-generation wireless communication, which promise a low-latency data transmission in a high user mobility scenario.
In particular, vehicle-to-infrastructure (V2I) communication \cite{kuutti2018survey} is an indispensable component of V2X networks. Thus, it is natural to deploy ISAC to support a high-data rate transmission as well as a high-resolution localization to facilitate V2I networks.
Indeed, various initial experimental results have shown that ISAC-based V2I networks can achieve a centimeter-order localization accuracy while maintaining a satisfactory communication rate \cite{wymeersch20175g}. Meanwhile, ISAC-based V2I networks not only conserve the spectral resources, but also reduce the hardware cost.
Therefore, research efforts towards the ISAC technology and the ISAC-based V2I systems have attracted tremendous attention from both academia and industry \cite{feng2020joint, xiao2020overview}.


Generally, according to whether the sensing targets can independently transmit sensing signals or not, ISAC can be divided into two types, i.e., (a) device-based ISAC and (b) device-free ISAC \cite{liu2021survey}.
One typical method of (a) is the integrated localization and communication (ILAC) \cite{xiao2020overview}.
For example, \cite{ghatak2018positioning} investigated an mmWave network for enabling both positioning and data-communication services.
In particular, by exploiting the native communication signal transmitted from the single-antenna users, the base station (BS) can provide the positioning service while satisfying heterogenous quality of service requirements.
Besides, the authors in \cite{destino2017trade} adopted a pilot-based beam training scheme for positioning in a single-user mmWave communication scenario.
In the proposed scheme, both the BS and the user are equipped with multi-antenna arrays to periodically transmit and receive the sensing signals for joint beam alignment, thus further improving the ISAC performance.
On top of \cite{destino2017trade}, \cite{kumar2018trade} extended the related work to a multi-user case, where the device-based multi-user single-input multiple-output (SIMO) uplink beam training scheme was explored.
By quantifying the localization-communication tradeoff as a function of beam training phase duration, the authors revealed that there exists an optimal beam training overhead to strike a balance between the effective rate and the localization accuracy.
Note that the above device-based methods focus on low-mobility users and require the sensing targets to independently transmit sensing signals to facilitate positioning at the BS.
However, for high-speed user scenarios, such as V2I networks, the use of sensing signal transmission would introduce a relatively long time delay.
This often leads to the outdatedness of location information acquisition jeopardizing to the beam tracking performance.

To overcome the above problems, dual-function radar-communication (DFRC) \cite{liu2020joint}, a typical example for device-free ISAC has been proposed, which enables a single device to provide dual functionalities: radar sensing and communication without the need of transmitting any sensing signals, such as pilot signals.
As early research efforts, the authors in \cite{roberton2003integrated} and \cite{saddik2007ultra} employed a classical radar waveform, i.e., chirp signals, as the carrier for information transmission and sensing. For example, the developed systems can transmit binary information bits via choosing whether to adopt the down-chirp waveform or the up-chirp waveform. Yet, this scheme can only support a quite low-speed data rate.
Recently, the orthogonal frequency division multiplexing (OFDM) technique has been introduced and exploited as a promising solution for the deployment of efficient DFRC \cite{xu2021wideband}. In contrast to conventional radar waveforms, OFDM-based radar signals can naturally decouple the Doppler and range estimators, which further improve the sensing performance \cite{sen2010adaptive, shi2017power}.
Besides the frequency domain, the rapid development of advanced MIMO techniques enable DFRC to explore the spatial domain to facilitate the implementation of ISAC.
For instance, pioneered by \cite{kumari2017ieee}, a MIMO radar was deployed for DFRC, where the pencil-like mainlobe was formed to facilitate target detection, meanwhile the information can be conveyed via adaptively controlling the powers of sidelobes. To further improve the overall performance of both radar and communication, \cite{liu2020jointtransmit} developed a joint transmit beamforming scheme to optimize the radar target detection accuracy under the constraint of quality of service in terms of communication.
On the other hand, to fully exploit the degrees-of-freedom brought by the spatial domain, the authors in \cite{huang2020majorcom} designed a carrier agile phased array radar-based DFRC scheme that adopts index modulation for information transmission.
Also, to provide a faster data rate and a higher sensing resolution for high mobility V2I networks, mmWave-based DFRC systems have recently been widely investigated, where a road side unit (RSU) is introduced to perform both beam tracking and motion parameter prediction to facilitate efficient ISAC.
Along with this line of thought, the authors in \cite{liu2020radar} developed an extended Kalman filtering (EKF) scheme to accurately track and predict the kinematic parameters of vehicles in mmWave frequency bands, so as to further improve the beam alignment performance. However, the proposed scheme in \cite{liu2020radar} incurs an exceedingly large computational complexity which is not suitable for practical implementation. To further simplify the prediction process, \cite{yuan2020bayesian} exploited the message passing technique to predict the motion parameters based on a Bayesian framework and the proposed scheme enjoys a relatively lower computational complexity compared with the EKF-based method, especially for large scale V2I networks.
It should be highlighted that although the aforementioned methods provided some intermediate solutions to the deployment of DFRC in mmWave-based V2I networks, these existing methods still adopt a cascaded two-stage scheme, i.e., predicting the future communication channels and then operating beam alignment, which is essentially a disjoint beamforming design for guaranteeing both the sensing and communication performance simultaneously.
Besides, it has been shown that the achievable beamforming performance is limited by the channel prediction accuracy.
Furthermore, the two-stage mechanism \cite{liu2020radar, yuan2020bayesian} directly aligns the transmit beam to the desired vehicle ignoring the existence of multiple access interference, which inevitably degrades the system sum-rate.
Thus, a practical predictive beamforming approach that can directly predict the joint beamforming matrix for next time slot without the need of explicit channel tracking/prediction is desired for the deployment of efficient ISAC in V2I networks.


Recently, the new developing deep learning (DL) technology has presented its powerful data-driven capability in various wireless communication applications \cite{wang2017deep}, ranging from channel estimation \cite{lxm2020deepresidual}, signal detection \cite{liu2020deeptransfer}, to resource allocation \cite{ye2019deep}. Note that the joint beamforming design in ISAC is generally a high-dimensional nonconvex problem \cite{liu2021survey}, thus it is challenging to find an optimal solution. On the other hand, optimizing the beamformer for ISAC is essentially an objective maximization problem which can be addressed in a data-driven approach to further improve the system performance.
In fact, the data-driven approach is a data-based processing mechanism where the DL technology is usually adopted to exploit effective features from the data to further improve the performance of a task \cite{wang2017deep, lecun2015deep}.
It has been proven that the data-driven DL approach has various advantages, such as model-free and non-linear mapping \cite{lecun2015deep}, which facilitates the development of efficient algorithms for addressing sophisticated optimization problems.
Motivated by this, in this paper, we adopt a DL approach to design a predictive joint beamforming scheme for V2I networks which not only maximizes the average achievable system sum-rate, but also guarantees the estimation accuracy of motion parameters for vehicles.
In fact, by exploiting the excellent feature extraction capability of DL \cite{liu2019deep, lecun2015deep}, the proposed predictive scheme can implicitly learn the features from historical estimated channels and directly predict the beamforming matrix to be used for the next time slot, which significantly reduces the system complexity.
The main contributions of our work are summarized as follows:
\begin{enumerate}[(1)]
\item We develop a predictive communication protocol, which bypasses the need of explicit channel tracking process to reduce the signaling overhead. Accordingly, a general predictive beamforming problem for ISAC systems is formulated to maximize the communication sum-rate while guaranteeing the sensing performance, where the Cramer-Rao lower bounds (CRLBs) are derived to characterize the estimation accuracy. Specifically, different from existing works \cite{liu2020radar, yuan2020bayesian} that over simplistically ignore the interference from other vehicles, our formulated ISAC beamforming problem takes into account the multiple access interference to design practical schemes for the implementation of ISAC systems.

\item To address the beamforming design problem, we propose a versatile unsupervised DL-based predictive beamforming framework, in which a penalty-based method is first exploited to transform the original optimization problem to an unconstrained optimization problem such that an unsupervised DL approach is then developed to handle the design problem in a data-driven manner.

\item As a realization of the proposed framework, a historical channels-based convolutional long short-term memory (LSTM) network (HCL-Net) is designed for predictive beamforming in ISAC-based V2I networks. Specifically, in HCL-Net, the historical estimated channels are considered as the input while the convolution and LSTM modules are adopted successively to exploit the spatial and temporal features of communication channels to further improve the learning capability for predictive beamforming.

\item We have conducted extensive simulations to verify the efficiency of the proposed algorithm in terms of both communication and sensing performance.
    In particular, the results demonstrate that the proposed predictive method can even guarantee harsh requirements on the sensing CRLBs and its achievable sum-rate approaches closely to the upper bound obtained by a genie-aided scheme, where the downlink channels in the latter are assumed to be parallel such that the optimal beamforming \cite{rao2003performance} can be derived with the availability of perfect instantaneous channel state information (ICSI).
\end{enumerate}

The remainder of this work is organized as follows.
The system model of the considered ISAC-based V2I network is introduced in Section \Rmnum{2}.
Section \Rmnum{3} formulates a general optimization problem for beamforming design in ISAC.
To tackle this problem, a DL-based predictive beamforming scheme for ISAC is proposed in Section \Rmnum{4}.
Then, we conduct simulations to verify the effectiveness of the proposed scheme in Section \Rmnum{5}.
Finally, Section \Rmnum{6} concludes this paper.

\emph{Notations}:
Unless otherwise specified, we adopt the bold uppercase letter, bold lowercase letter, and the normal font to represent the matrix, vector, and scalar, respectively.
Superscripts $T$, $*$, and $H$ denote the transpose operation, conjugate operation, and conjugate transpose operation, respectively.
$\mathbb{C}$ and $\mathbb{R}$ represent the sets of complex numbers and real numbers, respectively.
${\mathcal{CN}}( \bm{\mu},\mathbf{\Sigma} )$ and ${\mathcal{N}}( \bm{\mu},\mathbf{\Sigma} )$ denote the circularly symmetric complex Gaussian (CSCG) distribution and the real-valued Gaussian distribution, respectively, where $\bm{\mu}$ and $\mathbf{\Sigma}$ are the mean vector and the covariance matrix, respectively.
${\mathcal{U}}( a,b )$ denotes the uniform distribution within the range of $[a,b]$.
$\tb{0}$ denotes a zero vector or matrix according to requirements.
${\mathbf{I}}_N$ and ${\mathbf{1}}_N$ are used to represent the $N$-by-$N$ identity matrix and the $N$-by-$1$ all-ones vector, respectively.
$|\cdot|$ is the absolute value of a complex-valued number, $\|\cdot\|$ is the Euclidean norm of a vector, and $\|\cdot\|_F$ denotes the Frobenius norm of a matrix.
$(\cdot)^{-1}$ denotes the matrix inverse.
$\mathrm{diag}(\mathbf{x})$ denotes to generate a diagonal matrix based on vector $\mathbf{x}$.
$\mathbb{E}(\cdot)$ refers to the statistical expectation operation.
$\det(\cdot)$ denotes the determinant of a matrix.
$\frac{\partial f(x,y,\cdots)}{\partial x}$ denotes the partial derivative of a function $f(x,y,\cdots)$ with respect to variable $x$.
$\tb{A} \succeq \mathbf{B}$ denotes $\tb{A} - \mathbf{B}$ is positive semidefinite.
$\max(c,d)$ denotes the maximum between real-valued $c$ and $d$.
In addition, we adopt $\mathrm{Re}\{\cdot\}$ and $\mathrm{Im}\{\cdot\}$ to denote the real part and the imaginary part of a complex-valued matrix, respectively.

\vspace{-0.2cm}
\section{System Model}
\begin{figure}[t]
  \centering
  \includegraphics[width=\linewidth]{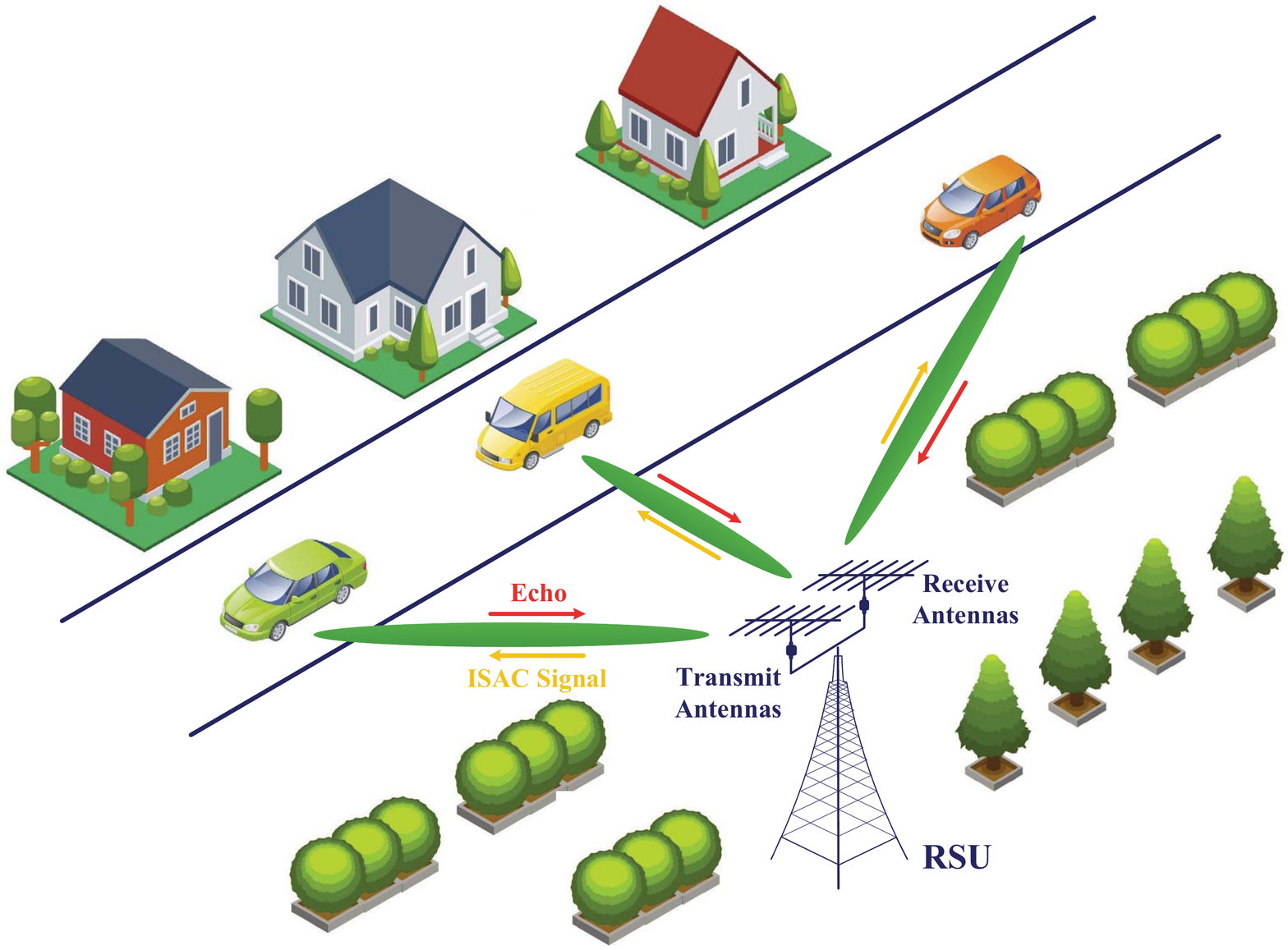}
  \caption{The considered ISAC-assisted V2I system with an RSU serving multiple vehicles.}\label{Fig:RSU_scenario}
\end{figure}

\vspace{-0.2cm}
As shown in Fig. \ref{Fig:RSU_scenario}, we consider an ISAC-assisted V2I network, where a roadside unit (RSU) serves $K$ single-antenna vehicles.
The RSU is a DFRC system which is equipped with an mmWave massive MIMO-type \cite{marzetta2016fundamentals} uniform linear array (ULA) consisting of $N_t$ transmit antennas and $N_r$ receive antennas. By exploiting full-duplex radio techniques \cite{barneto2021full} on transmit and receive antennas, the RSU can receive the potential signal echoes for sensing while maintaining uninterrupted downlink communications concurrently \cite{yuan2020bayesian}.

\subsection{Sensing Model}
Denote by $s_{k,n}(t)$ the ISAC downlink signal to be transmitted to the $k$-th, $k \in \{1,2,\cdots,K\}$, vehicle at time instant $t$ within the $n$-th, $n \in \{1,2,\cdots,N\}$, time slot.
The ISAC signal vector for all $K$ vehicles is given by $\tb{s}_{n}(t)=[s_{1,n}(t),s_{2,n}(t),\cdots,s_{K,n}(t) ]^T \in \mathbb{C}^{K \times 1}$ with $\mathbb{E}\{|s_{k,n}(t)|^2\}=1$.
Thus, the transmitted signal via the $N_t$ antennas at the RSU can be expressed as
\begin{equation}\label{Wsn_t}
\tilde{\tb{s}}_n(t) = \tb{W}_n \tb{s}_{n}(t) \in \mathbb{C}^{N_t \times 1},
\end{equation}
where $\tb{W}_n = [\tb{w}_{1,n},\tb{w}_{2,n},\cdots,\tb{w}_{K,n}]\in \mathbb{C}^{N_t\times K}$ denotes the downlink beamforming matrix at time slot $n$ and $\tb{w}_{k,n} \in \mathbb{C}^{N_t\times 1}$ is the dedicated beamforming vector for the $k$-th vehicle. In this case, the reflected signal echoes received at the RSU can be formulated as \cite{liu2020radar, yuan2020bayesian}
\begin{equation}\label{r_nt}
{\tb{r}}_n(t) =G \sum_{k=1}^K \beta_{k,n} e^{j2\pi \mu_{k,n}t} \tb{b}(\theta_{k,n}) \tb{a}^{\rm H}(\theta_{k,n})\tilde{\tb{s}}_{n}(t-\nu_{k,n}) + \tb{z}(t).
\end{equation}
Here, $G=\sqrt{N_t N_r}$ is the total antenna array gain and $\nu_{k,n}$ and $\mu_{k,n}$ are the time-delay and the Doppler frequency with respect to the $k$-th vehicle, respectively.
$\theta_{k,n}$ denotes the angle between the $k$-th vehicle and the RSU at time slot $n$.
$\beta_{k,n} = \frac{\varrho}{2d_{k,n}}$ is the reflection coefficient, where $\varrho$ denotes the fading coefficient based on the radar cross-section and $d_{k,n}$ is the distance between the $k$-th vehicle and the RSU at time slot $n$.
$\tb{z}(t)\in \mathbb{C}^{N_r\times 1}$ denotes the CSCG noise vector at the RSU.
In practical mmWave communication systems, a line-of-sight (LoS) channel model\footnotemark\footnotetext{The existence of occlusion or blocking obstacles may hinder the execution of the sensing and communication tasks \cite{liu2020joint}.
In addition, the echoes from the non-LoS channels can mislead the locations of the desired targets \cite{liu2020joint, liu2021survey}.
For ease of investigation, the study of these factors will be left for future work and only the LoS channel model is considered in this paper.} \cite{niu2015survey} is usually adopted and the transmit and the receive steering vectors at the RSU can be expressed as
\begin{equation}\label{}
  \tb{a}(\theta_{k,n})=\sqrt{\frac{1}{N_t}}[1,e^{-j\pi\cos\theta_{k,n}},\cdots,e^{-j\pi(N_t-1)\cos\theta_{k,n}}]^T
\end{equation}
and
\begin{equation}\label{}
  \tb{b}(\theta_{k,n})=\sqrt{\frac{1}{N_r}}[1,e^{-j\pi\cos\theta_{k,n}},\cdots,e^{-j\pi(N_r-1)\cos\theta_{k,n}}]^T,
\end{equation}
respectively.
Since a massive MIMO system is adopted at the RSU, the steering vectors with different angles are asymptotically orthogonal \cite{marzetta2016fundamentals}, i.e., $\forall k \neq k^{'}$, we have $|\tb{b}^H(\theta_{k,n})\tb{b}(\theta_{k^{'},n})| \approx 0$.
Thus, the inter-beam interference between different vehicles in the uplink echoes is negligible
and the RSU can distinguish different vehicles in terms of angle-of-arrivals (AoAs) for independent processing.
In this case, the received echo at the RSU from the $k$-th vehicle at time slot $n$, denoted by ${{r}}_{k,n}(t)$, can be extracted from (\ref{r_nt}) via a spatial filtering process \cite{van2004optimum}, i.e., multiplying an item of $\tb{b}^H(\dot{\theta}_{k,n})$ with ${\tb{r}}_n(t)$, which can be expressed as
\begin{equation}\label{r_kn}
\begin{aligned}
{{r}}_{k,n}(t) &= \tb{b}^H(\dot{\theta}_{k,n})\tb{r}_n(t) \\
 &= G \beta_{k,n} e^{j2\pi \mu_{k,n}t}
\tb{a}^{\rm H}(\theta_{k,n})\tilde{\tb{s}}_n(t-\nu_{k,n}) + {z}_{k,n}(t).
\end{aligned}
\end{equation}
Here, $\tb{b}^H(\dot{\theta}_{k,n})$ is the adopted receive beamforming vector for spatial filtering, $\dot{\theta}_{k,n}$ is an estimation of ${\theta}_{k,n}$ obtained via AoA estimation technology \cite{gross2015smart, van2004optimum}. For simplification, we assume $\dot{\theta}_{k,n} \approx  {\theta}_{k,n}$, i.e., $\tb{b}^H(\dot{\theta}_{k,n})\tb{b}({\theta}_{k,n}) \approx 1$.
In addition, ${z}_{k,n}(t) = \tb{b}^H(\dot{\theta}_{k,n})\tb{z}(t) \in \mathbb{C}$ is a CSCG noise variable, i.e., ${z}_{k,n}(t) \sim \mathcal{CN}(0,\sigma_z^2)$ and $\sigma_z^2$ is the noise variance.


\begin{figure}[t]
  \centering
  \includegraphics[width=\linewidth]{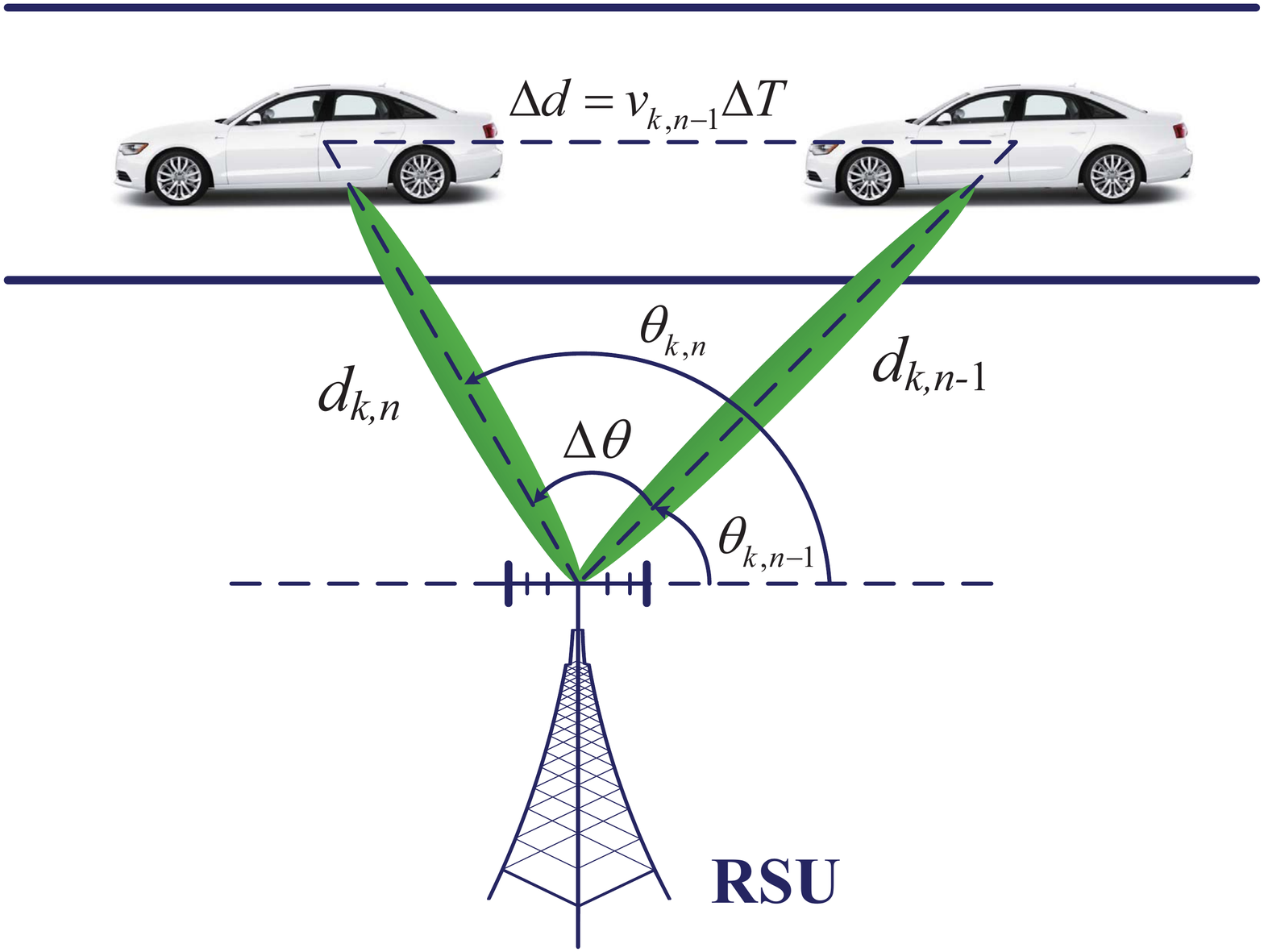}
  \caption{The kinematic model of vehicles in the considered V2I system.}\label{Fig:RSU_model}
\end{figure}

\begin{figure*}[t]
  \centering
  \includegraphics[width=\linewidth]{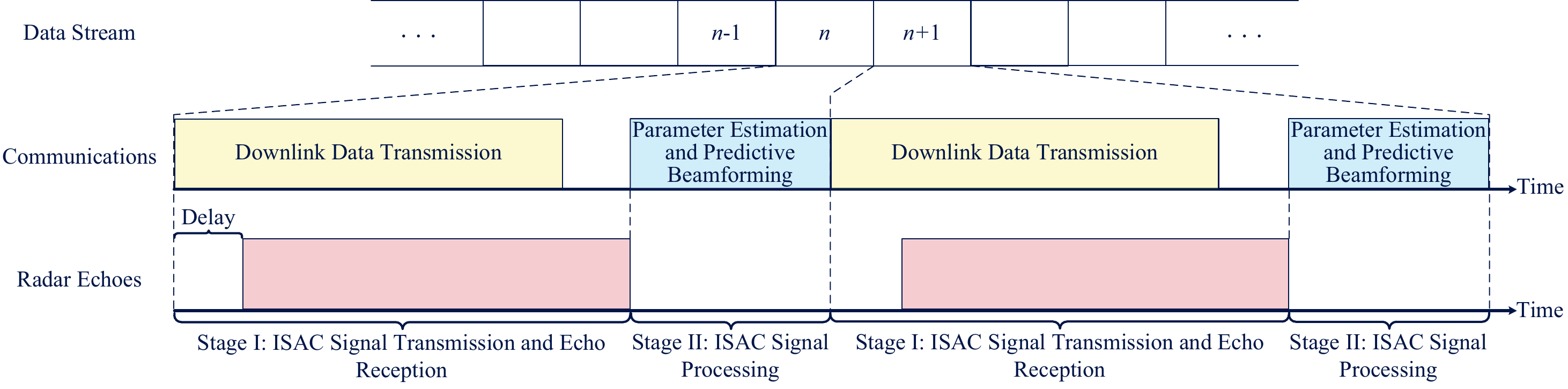}
  \caption{The developed communication protocol for ISAC in the considered V2I system.}\label{Fig:ISAC_frame_structure}
\end{figure*}

\subsection{Vehicle Mobility and Observation Model}
As shown in Fig. \ref{Fig:RSU_model}, the directions of all vehicles always keep in parallel to the road, i.e., the direction of velocity vector hardly changes \cite{wymeersch20175g}. Thus, we can characterize the velocity model in terms of the magnitude of the corresponding velocity vector, which is given by
\begin{equation}\label{}
  v_{k,n} = v_{k,n-1} + \Delta v_{k,n-1}, \forall k,n,
\end{equation}
where $v_{k,n}$ denotes the average velocity magnitude of the $k$-th vehicle at time slot $n$ and $\Delta v_{k,n-1}$ is the velocity increment within time slot $n-1$. For ease of study, we assume $v_{k,n} \sim \mathcal{U}(v_{\min},v_{\max})$, $\forall k,n$, with $v_{\min}$ and $v_{\max}$ being the minimum and the maximum of velocity magnitude, respectively \cite{wymeersch20175g}.
Based on this and the illustration in Fig. \ref{Fig:RSU_model}, the kinematic equations \cite{liu2020radar} characterizing the variations of the angle and the distance can be expressed as
\begin{equation}
\left\{
\begin{aligned}
  &\sin(\theta_{k,n} - \theta_{k,n-1})d_{k,n} = v_{k,n-1} \Delta T \sin \theta_{k,n-1},  \\
  &d_{k,n}^2 =  d_{k,n-1}^2 + (v_{k,n-1} \Delta T)^2 \\
  &\hspace{1cm} - 2d_{k,n-1}v_{k,n-1} \Delta T \cos\theta_{k,n-1},
\end{aligned}
\right.
\end{equation}
where $\Delta T$ is the time duration of each time slot.
In this case, the time-delay $\nu_{k,n}$ and the Doppler frequency $\mu_{k,n}$ can be estimated via the conventional matched-filtering method \cite{richards2014fundamentals}, i.e.,
\begin{equation}\label{matched-filtering}
\begin{aligned}
  &\{\tilde{\nu}_{k,n},\tilde{\mu}_{k,n}\}  \\
  &=  \arg \max_{\nu,\mu} \left|\int_0^{\Delta T_e}{{r}}_{k,n}(t)s_{k,n}^*(t- \nu)e^{-j2\pi\mu t } dt \right|^2,
  \end{aligned}
\end{equation}
where $\Delta T_e \leq \Delta T$ is the length of the received echo and $\tilde{\nu}_{k,n}$ and $\tilde{\mu}_{k,n}$ are the estimated values.
Based on the estimated values, we can then adopt interference cancellation scheme \cite{bechter2017analytical} to remove the multi-user interference in (\ref{r_kn}).
For simplification, we assume the interference can be removed ideally. Thus, after operating the interference cancellation in (\ref{r_kn}), we have ${{\dot{r}}}_{k,n}(t) = G \beta_{k,n} e^{j2\pi \mu_{k,n}t}
\tb{a}^{\rm H}(\theta_{k,n})\tb{w}_{k,n} {s}_{k,n}(t-\nu_{k,n}) + {z}_{k,n}(t)$.
In this case, the received echo at the RSU can be rewritten as an observation model of $\theta_{k,n}$, i.e.,
\begin{equation}\label{r_kn_theta}
\begin{aligned}
  \tilde{{r}}_{k,n} & \triangleq \int_0^{\Delta T_e}{{\dot{r}}}_{k,n}(t)s_{k,n}^*(t-\tilde{\nu}_{k,n})e^{-j2\pi\tilde{\mu}_{k,n} t } dt \\
   & = G \beta_{k,n}  \tb{a}^{\rm H}(\theta_{k,n})\mathbf{w}_{k,n} \int_0^{\Delta T_e}s_{k,n}(t-\nu_{k,n}) \\
   & \hspace{0.4cm}\cdot s_{k,n}^*(t-\tilde{\nu}_{k,n})e^{-j2\pi (\tilde{\mu}_{k,n}t - \mu_{k,n} t)} dt \\
   & \hspace{0.4cm} + \int_0^{\Delta T_e} z_{k,n}(t)s_{k,n}^*(t-\tilde{\nu}_{k,n})e^{-j2\pi\tilde{\mu}_{k,n} t} dt \\
& =  G \beta_{k,n} \xi  \tb{a}^{\rm H}(\theta_{k,n})\mathbf{w}_{k,n} + \tilde{{z}}_{k,n}.
\end{aligned}
\end{equation}
Here, $\xi \hspace{-0.1cm}=\hspace{-0.1cm} \int_0^{\Delta T_e}s_{k,n}(t-\nu_{k,n})s_{k,n}^*(t-\tilde{\nu}_{k,n})e^{-j2\pi (\tilde{\mu}_{k,n}t - \mu_{k,n} t)} dt$ is the matched-filtering gain.
$\tilde{{z}}_{k,n} \sim \mathcal{CN}(0,\sigma_{r_k}^2)$ is the noise term with $\sigma_{r_k}^2$ being the noise variance.
Accordingly, the terms $d_{k,n}$ and $v_{k,n}$ obey the observation models, which yields
\begin{equation}\label{tau_kn_d}
  \tilde{\nu}_{k,n} = {\nu}_{k,n} + \epsilon_{k,n} = \frac{2d_{k,n}}{c} + \epsilon_{k,n}
\end{equation}
and
\begin{equation}\label{mu_kn_v}
  \tilde{\mu}_{k,n} = {\mu}_{k,n} + \varepsilon_{k,n} = \frac{2\dot{v}_{k,n} f_c}{c} + \varepsilon_{k,n},
\end{equation}
respectively. Here, $f_c$ is the carrier frequency, $c$ is the speed of signal propagation, $\dot{v}_{k,n}$ denotes the radial velocity. $\epsilon_{k,n} \sim \mathcal{N}(0,\sigma_{\nu_k}^2)$ \cite{tse2005fundamentals} with the noise variance $\sigma_{\nu_k}^2$ and $\varepsilon_{k,n} \sim \mathcal{N}(0,\sigma_{\mu_k}^2)$ with the noise variance $\sigma_{\mu_k}^2$ are the estimation errors of ${\nu}_{k,n}$ and ${\mu}_{k,n}$, respectively\footnotemark\footnotetext{Note that the noise terms in (\ref{tau_kn_d}) and (\ref{mu_kn_v}) are introduced by the matched-filtering process in (\ref{matched-filtering}), whose distributions mainly depend on the noise term in (\ref{matched-filtering}).
In fact, the noise term in (\ref{matched-filtering}) has only one random component, i.e., a CSCG vector $\tb{z}(t)$ as defined in (\ref{r_nt}). In this case, the noise term in (\ref{matched-filtering}) is a CSCG variable. Thus, the noise terms in (\ref{tau_kn_d}) and (\ref{mu_kn_v}) also follow the Gaussian distribution.}.
Note that $\sigma_{\nu_k}^2$ and $\sigma_{\mu_k}^2$ generally depend on the signal-to-noise ratio ($\mathrm{SNR}$) at the RSU \cite{kay1993fundamentals}, i.e., the received SNR of (\ref{r_kn}), which is derived as
\begin{equation}\label{SNR}
\begin{aligned}
   &\mathrm{SNR}_{k,n}\\
   &= \frac{\mathbb{E}\left\{\left|G \beta_{k,n} e^{j2\pi \mu_{k,n}t}
\tb{a}^{\rm H}(\theta_{k,n}) \mathbf{w}_{k,n} {s}_{k,n}(t-\nu_{k,n}) \right|^2\right\}}{\mathbb{E}\left\{\left|\sum_{i \neq k}^K G \beta_{i,n} \tb{a}^{\rm H}(\theta_{k,n}) \mathbf{w}_{i,n}{s}_{i,n}(t-\nu_{k,n}) + {z}_{k,n}(t)\right|^2\right\}} \\
&= \frac{ G^2|\beta_{k,n}|^2|\mathbf{a}^H(\theta_{k,n})\mathbf{w}_{k,n}|^2}{\sum_{i \neq k}^K G^2|\beta_{i,n}|^2 \left|\tb{a}^{\rm H}(\theta_{k,n}) \mathbf{w}_{i,n}\right|^2 + \sigma_z^2}, \end{aligned}
\end{equation}
where the terms of interference from other users are considered as noise, $\mathbb{E}\{|s_{k,n}(t)|^2\}=1$, and $\mathbb{E}\{|z_{k,n}(t)|^2\}=\sigma_z^2$, as defined in Section II-A.
Based on this, we have \cite{liu2020radar}
\begin{equation}\label{sigma_tau}
  \sigma_{\nu_k}^2 = \frac{\rho_{\nu}^2(\sum_{i \neq k}^K G^2|\beta_{i,n}|^2\left|\tb{a}^{\rm H}(\theta_{k,n}) \mathbf{w}_{i,n}\right|^2+\sigma_z^2)}{ G^2|\beta_{k,n}|^2|\mathbf{a}^H(\theta_{k,n})\mathbf{w}_{k,n}|^2}
\end{equation}
and
\begin{equation}
  \sigma_{\mu_k}^2 = \frac{\rho_{\mu}^2(\sum_{i \neq k}^K G^2|\beta_{i,n}|^2 \left|\tb{a}^{\rm H}(\theta_{k,n}) \mathbf{w}_{i,n}\right|^2+\sigma_z^2)}{ G^2|\beta_{k,n}|^2|\mathbf{a}^H(\theta_{k,n})\mathbf{w}_{k,n}|^2},
\end{equation}
respectively, where $\rho_{\nu}$ and $\rho_{\mu}$ are given constants depending on the detailed system deployment and estimation algorithms.
In particular, it is shown that noise variances highly depend on the downlink beamforming vector via term $|\mathbf{a}^H(\theta_{k,n})\mathbf{w}_{k,n}|$, thus we can improve the observation accuracy through adjusting $\mathbf{w}_{k,n}$.

\subsection{Communication Model}
At time instant $t$ within time slot $n$, the $k$-th vehicle receives the downlink signal from the RSU, which can be expressed as
\begin{equation}\label{cmodel}
\vartheta_{k,n}(t) = \tilde{G} \sqrt{\alpha_{k,n}} e^{j2\pi \mu_{k,n}t}\tb{a}^{ H}(\theta_{k,n})\sum_{i=1}^K \mathbf{w}_{i,n}{s}_{i,n}(t) + \eta_{k,n}(t).
\end{equation}
Here, $\tilde{G} = \sqrt{N_t}$ is the antenna gain,  $\alpha_{k,n} = \alpha_0(d_{k,n}/d_0)^{-\zeta}$ is the path loss coefficient, where $\alpha_0$ is the path loss at reference distance $d_0$, and $\zeta$ is the associated path loss exponent. In addition, $\eta_{k,n}(t) \sim \mathcal{CN}(0,\sigma_k^2)$ denotes the noise at the $k$-th vehicle with $\sigma_k^2$ being the noise variance.
For each vehicle receiver, the interference originates from the signals transmitted for other vehicles, which should be taken into account in beamforming design to further improve the sum-rate performance.
It should be highlighted that the effect of the multi-user interference has been considered during both the sensing and communication process since both (\ref{r_kn}) and (\ref{cmodel}) have the multi-user interference component, i.e., $\sum_{i\neq k}^K \mathbf{w}_{i,n}{s}_{i,n}(t)$.
In particular, the received signal-to-interference-plus-noise ratio (SINR) at the $k$-th vehicle within time slot $n$ can be expressed as
\begin{equation}\label{SINR}
\begin{aligned}
  \mathrm{SINR}_{k,n}(\mathbf{h}_{k,n},\mathbf{w}_{k,n}) &= \frac{\left|\tilde{G} \sqrt{\alpha_{k,n}}  \tb{a}^{H}(\theta_{k,n})\mathbf{w}_{k,n}\right|^2 }
  { \sum\limits_{k^{'} \neq k}^K \left|\tilde{G} \sqrt{\alpha_{k,n}}  \tb{a}^{H}(\theta_{k,n}) \mathbf{w}_{k^{'},n}\right|^2 + \sigma_{k}^2} \\
&= \frac{ \left| \mathbf{h}_{k,n}^H\mathbf{w}_{k,n}\right|^2 }{\sum\limits_{k^{'} \neq k}^K \left| \mathbf{h}_{k,n}^H\mathbf{w}_{k^{'},n}\right|^2 + \sigma_{k}^2},
\end{aligned}
\end{equation}
where $\mathbf{h}_{k,n}^H = \tilde{G}\sqrt{\alpha_0(d_{k,n}/d_0)^{-\zeta}}\tb{a}^{H}(\theta_{k,n})$ is the equivalent channel vector between the $k$-th vehicle and the RSU at time slot $n$.

\subsection{Proposed Protocol}
Note that if the beamforming matrix $\mathbf{W}_n$ was optimized in time slot $n-1$, the expected communication performance at time slot $n$ can be guaranteed in advance.
Inspired by this, we develop a hierarchical transmission protocol for ISAC in the considered V2I system, where a predictive beamforming matrix for the next time slot is preset in advance, thus bypassing the real-time channel tracking or motion parameter prediction to further reduce the signaling overhead compared with existing works, e.g., \cite{kumari2017ieee, liu2020jointtransmit, huang2020majorcom, liu2020radar, yuan2020bayesian}.
As depicted in Fig. \ref{Fig:ISAC_frame_structure}, the data stream is divided into different time slots. In each time slot, the RSU requires to send ISAC signals for both the downlink communication and the sensing tasks.
Specifically, there are two stages for each time slot in the developed protocol, i.e., Stage I: ISAC signal transmission and echo reception; Stage II: ISAC signal processing.
Taking the $n$-th time slot for example, in Stage I, the RSU transmits ISAC signals with the optimized beamforming matrix obtained from the last prediction and receives echoes. In Stage II, the RSU first estimates vehicles' motion parameters at the current time slot, i.e., time slot $n$, based on the received echoes from different vehicles and then optimizes the beamforming matrix for the next time slot, i.e., time slot $n+1$, according to the current and historical estimated channels. As such, the RSU can bypass the process of acquiring the ICSI and directly design the beamforming matrix to guarantee the ISAC performance. In this paper, we mainly focus on the predictive beamforming design\footnotemark\footnotetext{Note that as described in Fig. \ref{Fig:RSU_scenario}, Stage I can be handled via some full-duplex radio techniques, e.g., \cite{barneto2021full}.} in Stage II and the related problem formulation will be given in the next section.

\begin{figure}[t]
  \centering
  \includegraphics[width=\linewidth]{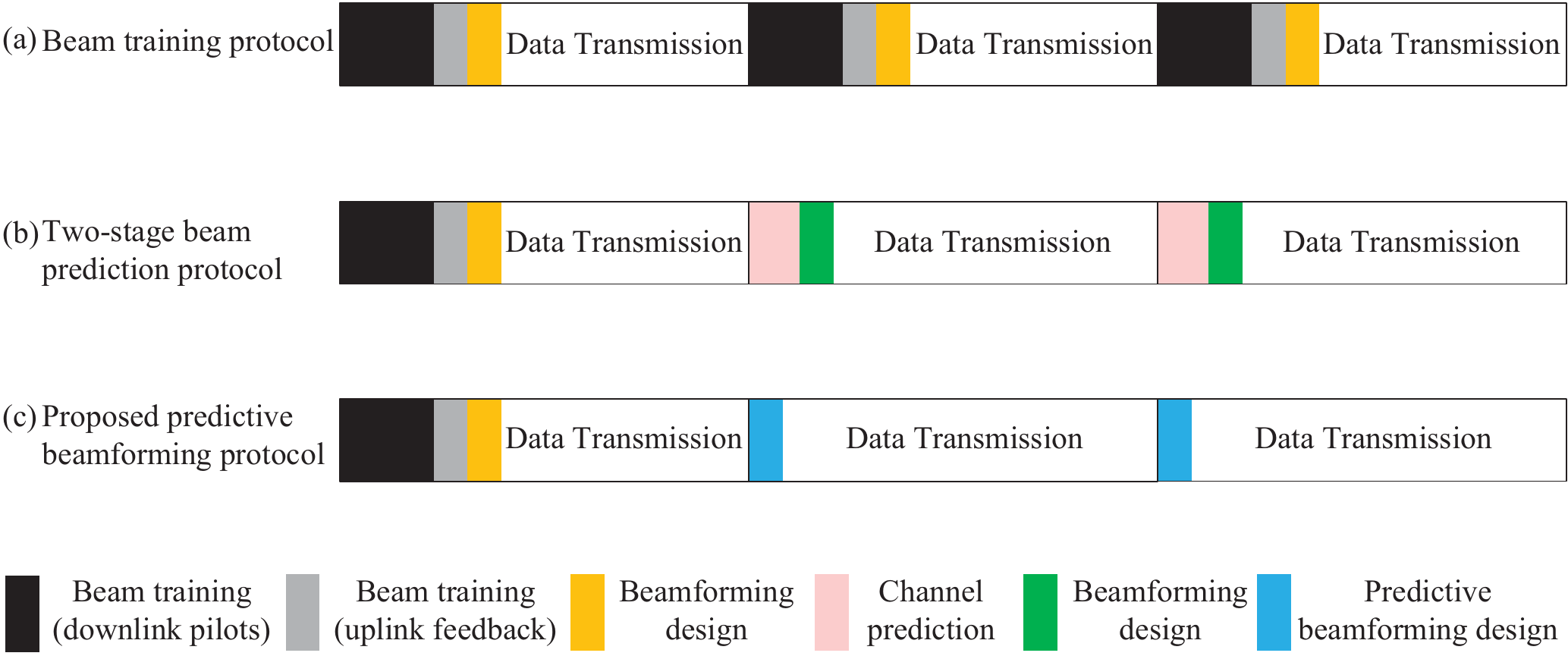}
  \caption{Frameworks of different beamforming protocols for ISAC.}\label{Fig:BF_framework_comparison}
\end{figure}

In addition, a comparison in terms of the complexity of the frameworks with different existing beamforming protocols are shown in Fig. \ref{Fig:BF_framework_comparison}, where three beamforming protocols are studied: (a) beam training protocol \cite{zhang2019codebook}, (b) two-stage beam prediction protocol \cite{liu2020radar}, and (c) the proposed predictive beamforming protocol.
It can be seen that both the downlink pilots and the uplink feedback are indispensable for beam training in (a), which introduces huge complexity and signaling overhead.
Although (b) adopts a DFRC block which does not require beam training, it still needs a channel prediction-beamforming two-stage disjoint process, which also brings considerable computational overhead.
On top of (b), our proposed (c) develops a learning-based joint predictive mechanism to directly predict the beamforming matrix based on the historical channels without the need of performing explicit channel prediction and beam training.
Thus, our proposed predictive beamforming protocol has a lower complexity compared with the other related protocols.

\section{Problem Formulation}
In this paper, we aim to maximize the average achievable sum-rate of the V2I downlink communication while guaranteeing the sensing performance for vehicles. In the following, we will first derive CRLBs of motion parameter estimation to quantify the sensing performance and then formulate the corresponding optimization problem.

\subsection{Cramer-Rao Lower Bound for Parameter Estimation}
In this section, we derive the Cramer-Rao lower bound (CRLB) to characterize the estimation accuracy. To start with, denote by $\mathbf{y}_{k,n} = [\tilde{{r}}_{k,n}, \tilde{\nu}_{k,n}, \tilde{\mu}_{k,n}]^T \in \mathbb{C}^{(N_r + 2) \times 1}$ and $ \mathbf{x}_{k,n} = [\theta_{k,n},d_{k,n},\dot{v}_{k,n}]^T$ the observation vector and the motion parameter vector, respectively, we have
\begin{equation}\label{}
  \tb{y}_{k,n} = \mathbf{g}(\tb{x}_{k,n}) + \tb{u}_{k,n},
\end{equation}
where $\mathbf{g}(\cdot)$ is defined according to (\ref{r_kn_theta})-(\ref{mu_kn_v}), $\tb{u}_{k,n} = [\tilde{{z}}_{k,n},  \epsilon_{k,n}, \varepsilon_{k,n}]^T$, and $\tb{y}_{k,n} \sim \mathcal{CN}(g(\tb{x}_{k,n}),\mathbf{\Sigma} )$ with $\mathbf{\Sigma} = \mathrm{diag} ([\sigma_{r_k}^2, \sigma_{\nu_k}^2, \sigma_{\mu_k}^2]) $ being the covariance matrix of $\tb{y}_{k,n}$\footnotemark\footnotetext{Note that the correlations among noise terms can provide additional observation information of the desired sensing targets, which can be exploited to further improve the sensing performance.
In this paper, we assume the noise terms are uncorrelated since this assumption represents an unfavourable scenario of ISAC systems \cite{liu2021survey, liu2020radar} and the adopted framework can serve as a solid foundation for possible further extensions.}. In this case, the conditional probability density function (PDF) of $\mathbf{y}_{k,n}$ given $\mathbf{x}_{k,n}$ can be expressed as
\begin{equation}\label{}
\begin{aligned}
  &p(\mathbf{y}_{k,n}|\mathbf{x}_{k,n}) = \frac{1}{\pi^{N_r+2}\det(\mathbf{\Sigma})}\\
  &\hspace{1.0cm}\exp\left(-(\mathbf{y}_{k,n} - \mathbf{g}(\mathbf{x}_{k,n})^H)\mathbf{\Sigma}^{-1}(\mathbf{y}_{k,n} - \mathbf{g}(\mathbf{x}_{k,n}))\right).
\end{aligned}
\end{equation}
According to the CRLB theorem \cite{kay1993fundamentals}, the Fisher information matrix (FIM) \cite[p. 529]{kay1993fundamentals} for $\tb{x}_{k,n}$ is given by (\ref{FIM_equation}), as shown at the top of the next page, where $\tb{F}(\tb{x}_{k,n})$ denotes the FIM of $\tb{x}_{k,n}$, $\mathbb{E}\left[(\tb{y}_{k,n} - \tb{g}(\tb{x}_{k,n}))(\tb{y}_{k,n} - \tb{g}(\tb{x}_{k,n}))^H\right] = \mathbf{\Sigma}$ and
\newcounter{mytempeqncnt1}
\begin{figure*}[!t]
\normalsize
\setcounter{mytempeqncnt1}{\value{equation}}
\setcounter{equation}{18}
{\begin{equation}\label{FIM_equation}
\begin{aligned}
  \tb{F}(\tb{x}_{k,n}) & = \mathbb{E}\left[\frac{\partial \ln p(\mathbf{y}_{k,n}|\mathbf{x}_{k,n}) }{\partial \mathbf{x}_{k,n}^*} \left(\frac{\partial \ln p(\mathbf{y}_{k,n}|\mathbf{x}_{k,n}) }{\partial \mathbf{x}_{k,n}^*}\right)^H \right]  \\
  & = \mathbb{E}\left[\left(\frac{\partial \tb{g}(\tb{x}_{k,n}) }{\partial \tb{x}_{k,n}}\right)^H\mathbf{\Sigma}^{-1}(\tb{y}_{k,n} - \tb{g}(\tb{x}_{k,n}))(\tb{y}_{k,n} - \tb{g}(\tb{x}_{k,n}))^H \mathbf{\Sigma}^{-1}\frac{\partial \tb{g}(\mathbf{x}_{k,n})}{\partial \tb{x}_{k,n}} \right]  \\
  & = \left(\frac{\partial \tb{g}(\tb{x}_{k,n}) }{\partial \tb{x}_{k,n}}\right)^H\mathbf{\Sigma}^{-1}\mathbb{E}\left[(\tb{y}_{k,n} - \tb{g}(\tb{x}_{k,n}))(\tb{y}_{k,n} - \tb{g}(\tb{x}_{k,n}))^H\right]\mathbf{\Sigma}^{-1}\frac{\partial \tb{g}(\mathbf{x}_{k,n})}{\partial \tb{x}_{k,n}}  \\
   & = \left(\frac{\partial \tb{g}(\tb{x}_{k,n})}{\partial \mathbf{x}_{k,n}}\right)^H\mathbf{\Sigma}^{-1}\frac{\partial \tb{g}(\tb{x}_{k,n})}{\partial \mathbf{x}_{k,n}}.
   \end{aligned}
\end{equation}}
\setcounter{equation}{\value{mytempeqncnt1}}
\hrulefill
\vspace*{4pt}
\end{figure*}
\setcounter{equation}{19}
\begin{equation}\label{P}
  \frac{\partial \tb{g}(\mathbf{x}_{k,n})}{\partial \mathbf{x}_{k,n}} = \left[
  \begin{matrix}
  \frac{\partial \tilde{{r}}_{k,n}}{\partial \theta_{k,n}}      & 0      & 0      \\
  0      & \frac{2}{c}       & 0      \\
  0      & 0      & \frac{2f_c}{c}      \\
  \end{matrix}
  \right] \in \mathbb{C}^{(N_r+2) \times (N_r+2)}.
\end{equation}
Based on this, the mean squared error (MSE) matrix of $\tb{x}_{k,n}$ is bounded by below:
\begin{equation}\label{}
  \mathbb{E}\left[(\tilde{\tb{x}}_{k,n} - \tb{x}_{k,n})(\tilde{\tb{x}}_{k,n} - \tb{x}_{k,n})^H\right] \succeq \tb{F}^{-1}(\tb{x}_{k,n}).
\end{equation}
Accordingly, the MSE lower bounds of $\theta_{k,n}$ and $d_{k,n}$ are given by
\begin{equation}\label{}
  \mathbb{E}\left[(\tilde{\theta}_{k,n} - \theta_{k,n})^2\right] \geq f_{11} \triangleq \mathrm{CRLB}(\theta_{k,n},\tb{w}_{k,n})
\end{equation}
and
\begin{equation}\label{}
  \mathbb{E}\left[(\tilde{d}_{k,n} - d_{k,n})^2\right] \geq f_{22} \triangleq \mathrm{CRLB}(d_{k,n},\tb{w}_{k,n}) ,
\end{equation}
respectively. Here, $f_{ij}$ denotes the $i$-th row and the $j$-th column element of $\tb{F}^{-1}(\tb{x}_{k,n})$ and we have
\begin{equation}\label{CRLB_theta}
  \mathrm{CRLB}(\theta_{k,n},\tb{w}_{k,n}) =  \left[\frac{1}{\sigma_{r_k}^2}\left(\frac{\partial \tilde{{r}}_{k,n}}{\partial \theta_{k,n}} \right)^H  \frac{\partial \tilde{{r}}_{k,n}}{\partial \theta_{k,n}}\right]^{-1}
\end{equation}
with
\begin{equation}\label{partial_r}
\begin{aligned}
  &\frac{\partial \tilde{{r}}_{k,n}}{\partial \theta_{k,n}} \\
  & = -\sqrt{N_r}\beta_{k,n}\xi\hspace{-0.05cm}\sum_{n_t=2}^{N_t} w_{k,n}^{(n_t)} e^{j\pi(n_t-1)\cos\theta_{k,n}} j\pi(n_t \hspace{-0.05cm} - \hspace{-0.05cm} 1)\sin\theta_{k,n}
\end{aligned}
\end{equation}
and
\begin{equation}\label{CRLB_d}
  \mathrm{CRLB}(d_{k,n},\tb{w}_{k,n}) =
  \left[\frac{1}{\sigma_{\nu_k}^2}\left(\frac{2}{c}\right)^2\right]^{-1},
\end{equation}
where $w_{k,n}^{(n_t)}$ denotes the $n_t$-th, $n_t \in \{1,2,\cdots, N_t\}$, element of $\tb{w}_{k,n}$, $z_{k,n}^{(n_r)}$ denotes the $n_r$-th, $n_r \in \{1,2,\cdots, N_r\}$, element of $\dot{\tb{z}}_{k,n}$,  and terms $\sigma_{r_k}^2$ and $\sigma_{\nu_k}^2$ are defined in (\ref{r_kn_theta}) and (\ref{sigma_tau}), respectively.
In addition, $\mathrm{CRLB}(\theta_{k,n},\tb{w}_{k,n})$ and $\mathrm{CRLB}(d_{k,n},\tb{w}_{k,n})$ represent the CRLBs of estimating $\theta_{k,n}$ and $d_{k,n}$ given $\tb{w}_{k,n}$, respectively.

\textbf{Remark 1}: Note that $\tilde{{r}}_{k,n}$ in (\ref{CRLB_theta}) and $\sigma_{\nu_k}^2$ in (\ref{CRLB_d}) are the functions of $\tb{w}_{k,n}$, as defined in (\ref{r_kn_theta}) and (\ref{sigma_tau}), respectively.
Through optimizing the beamforming matrix $\mathbf{W}_n$, one may consider aligning each transmit beam to the desired vehicle for accurate sensing by improving the desired signal strength, as commonly adopted in e.g., \cite{liu2020radar, yuan2020bayesian}. Yet, it ignores the impacts created by multiple access interference on the downlink communication. Thus, a proper design of beamforming to strike a balance between sensing performance and communication performance is needed.

\subsection{Problem Formulation}
Based on the derived CRLBs of motion parameter estimation, we aim to maximize the average achievable sum-rate via optimizing the beamforming matrix at the RSU subject to the maximum transmit power constraint and the CRLB constraints of the sensing task.
Thus, the problem formulation at time slot $n$ can be expressed as
\begin{align}
(\mathrm{P}1): &\mathop{\max}\limits_{{\mathbf{W}}_n } ~ \mathbb{E}_{\mathbf{H}_{n}|\tb{\Omega}_{n}^{\tau}}
\left\{\sum_{k = 1}^K\mathrm{log}_2\left( 1 + \mathrm{SINR}_{k,n}(\mathbf{h}_{k,n},\mathbf{w}_{k,n})\right) \hspace{-0.05cm} \right\}  \label{P1_OP} \\
&\mathrm{s.t.}~
\mathbb{E}_{\mathbf{c}_{n}|\mathbf{\Theta}_{n}^{\tau}} \left\{\frac{1}{K}\sum_{k=1}^K\mathrm{CRLB}(\theta_{k,n},\mathbf{w}_{k,n})\right\} \leq \gamma_{\theta}, \label{P1_C_theta} \\
&\hspace{0.6cm}
\mathbb{E}_{\mathbf{d}_{n}|\mathbf{D}_{n}^{\tau}} \left\{\frac{1}{K}\sum_{k=1}^K\mathrm{CRLB}(d_{k,n},\mathbf{w}_{k,n})\right\} \leq \gamma_d, \label{P1_C_d} \\
&\hspace{0.6cm} \|\mathbf{W}_n\|_F^2\leq P. \label{P1_power}
\end{align}
Here, $\tb{W}_n$ defined in (\ref{Wsn_t}) and $\tb{h}_{k,n}$ defined in (\ref{SINR}) are the beamforming matrix and the channel vector, respectively. $\mathbb{E}_{\mathbf{H}_{n}|\tb{\Omega}_{n}^{\tau}}\{\cdot\}$ in the objective function is the ergodic average with respect to (w.r.t.) $\tb{H}_n = [\tb{h}_{1,n},\tb{h}_{2,n},\cdots,\tb{h}_{K,n}]$, given the historical estimated channels $\tb{\Omega}_{n}^{\tau} \triangleq [\tilde{\mathbf{H}}_{n-1},\tilde{\mathbf{H}}_{n-2},\cdots,\tilde{\mathbf{H}}_{n-\tau}]$,
where $\tilde{\mathbf{H}}_{n} =[\tilde{\mathbf{h}}_{1,n},\tilde{\mathbf{h}}_{2,n},\cdots,\tilde{\mathbf{h}}_{K,n}]$ and $\tilde{\mathbf{h}}_{k,n} = \tilde{G}\sqrt{\alpha_0 \tilde{d}_{k,n}^{-\zeta}}\tb{a}^{ H}(\tilde{\theta}_{k,n})$ with $\tilde{\theta}_{k,n}$ and $\tilde{d}_{k,n}$ being the estimated angles and distances, respectively.
Note that the term $\mathbf{\Omega}_{n}^{\tau}$ represents the set of historical estimated channels, which is related to the number of transmit antennas, the path loss, the estimated angles and distances.
Similarly, $\mathbb{E}_{\mathbf{c}_{n}|\mathbf{\Theta}_{n}^{\tau}}$ is the ergodic average w.r.t. $\tb{c}_n = [\theta_{1,n},\theta_{2,n},\cdots,\theta_{K,n}]^T$, given the historical estimated angles $\mathbf{\Theta}_{n}^{\tau} \triangleq [\tilde{\mathbf{c}}_{n-1},\tilde{\mathbf{c}}_{n-2},\cdots,\tilde{\mathbf{c}}_{n-\tau}]$
with $\tilde{\mathbf{c}}_{n} =[\tilde{\theta}_{1,n},\tilde{\theta}_{2,n},\cdots,\tilde{\theta}_{K,n}]^T$.
In addition, the expectation $\mathbb{E}_{\mathbf{d}_{n}|\mathbf{D}_{n}^{\tau}}$ is taken w.r.t. $\tb{d}_n = [d_{1,n},d_{2,n},\cdots,d_{K,n}]^T$, given the historical estimated distances $\mathbf{D}_{n}^{\tau} \triangleq [\tilde{\mathbf{d}}_{n-1},\tilde{\mathbf{d}}_{n-2},\cdots,\tilde{\mathbf{d}}_{n-\tau}]$
with $\tilde{\mathbf{d}}_{n} =[\tilde{d}_{1,n},\tilde{d}_{2,n},\cdots,\tilde{d}_{K,n}]^T$.
On the other hand, since only the historical channel information (i.e., from time slot $n-1$ to time slot $n-\tau$) can be exploited when designing the beamforming matrix for time slot $n$, the ergodic average is adopted to characterize the performance of communication and sensing tasks at time slot $n$ \cite{tse2005fundamentals}.
Moreover, $\gamma_{\theta}$ and $\gamma_d$ are the maximum tolerable CRLB thresholds to guarantee the sensing performance. $P$ in (\ref{P1_power}) is the power budget at the RSU for every time slot $n$.

According to Jensen's inequality \cite{tse2005fundamentals}, it can be derived that if one aims to improve the sum-rate, $\tb{W}_n$ should be designed to make all vehicles maintain the same received SINRs as defined in (\ref{SINR}).
On the other hand, if we aim to minimize the CRLB bound,
we should adapt $\tb{W}_n$ such that all the received SNR values at the RSU from different vehicles in (\ref{SNR}) are identical as possible.
Thus, there exists a tradeoff between communication rate and sensing accuracy, which will be verified via simulations in Section V.
Generally, problem ($\mathrm{P}1$) is challenging since it is intractable to derive the closed-forms of (\ref{P1_OP}), (\ref{P1_C_theta}), and (\ref{P1_C_d}).
Besides, even if the perfect CSI is available, the objective function and constraints are non-convex w.r.t. $\mathbf{w}_{k,n}$.
As an alternative, we will adopt a data-driven approach to develop a learning-based predictive beamforming scheme to address problem ($\mathrm{P}1$) in the following.


\section{Deep Learning-based Predictive Beamforming for ISAC}

Note that generally the DL approach is designed to handle unconstrained optimization problems \cite{goodfellow2016deep}.
To facilitate the application of the DL approach, we exploit the penalty method which transforms the constrained optimization problem (P1) to an unconstrained problem equivalently \cite{gill2019practical}.
In the following, we will first introduce the developed DL-based predictive beamforming framework and then design the historical channels-based convolutional LSTM network (HCL-Net) structure as a realization of the developed framework. Finally, we will propose the HCL-Net-based predictive beamforming algorithm.

\subsection{DL-based Predictive Beamforming Framework for ISAC}
\begin{figure}[t]
  \centering
  \includegraphics[width=\linewidth]{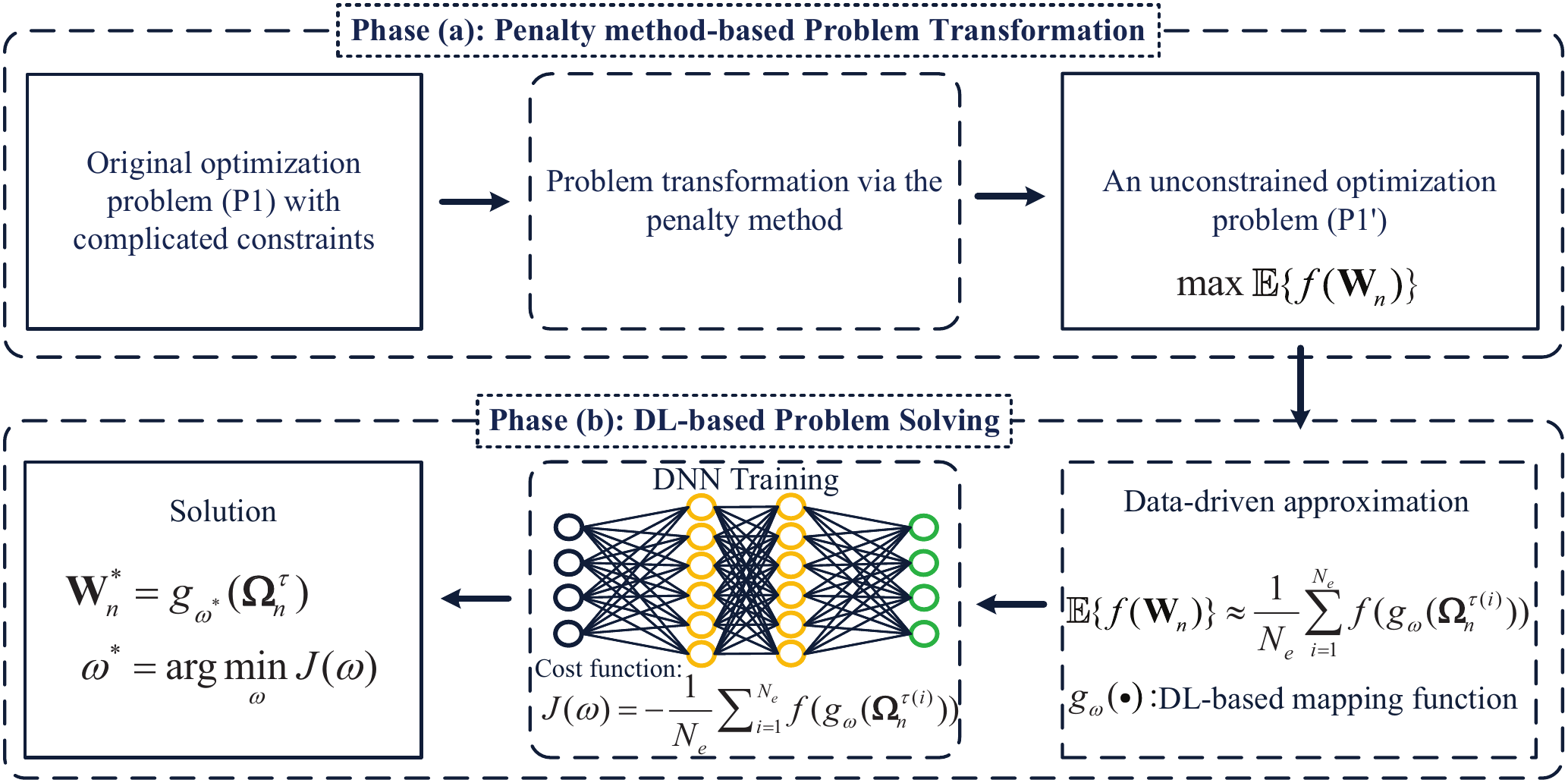}
  \caption{The developed predictive beamforming framework for ISAC.}\label{Fig:PBF_framework}
\end{figure}

First, we handle the non-convex constraints. According to \cite{gill2019practical}, the penalty method can be adopted to transform the constrained optimization problem (P1) to an equivalent unconstrained problem, which is given by
\begin{equation}\label{P1'}
\begin{aligned}
&(\mathrm{P}1^{'}): \mathop{\max}_{{\mathbf{W}}_n } ~ \mathbb{E}_{\mathbf{H}_{n}|\tb{\Omega}_{n}^{\tau}}
\left\{\sum_{k = 1}^K\mathrm{log}_2\left( 1 + \mathrm{SINR}_{k,n}(\mathbf{h}_{k,n},\mathbf{w}_{k,n})  \right) \hspace{-0.1cm} \right\}  \\
& - \hspace{-0.1cm}\lambda_1\hspace{-0.1cm}\left[\max\left(0, \mathbb{E}_{\mathbf{c}_{n}|\mathbf{\Theta}_{n}^{\tau}} \left\{\frac{1}{K}\sum_{k=1}^K\mathrm{CRLB}(\theta_{k,n},\mathbf{w}_{k,n})\right\} - \gamma_{\theta} \hspace{-0.05cm}\right)\hspace{-0.05cm}\right]^2 \\
& - \hspace{-0.1cm}\lambda_2\hspace{-0.1cm}\left[\max\left(0,\mathbb{E}_{\mathbf{d}_{n}|\mathbf{D}_{n}^{\tau}} \left\{\frac{1}{K}\sum_{k=1}^K\mathrm{CRLB}(d_{k,n},\mathbf{w}_{k,n})\right\} - \gamma_d \hspace{-0.05cm}\right)\hspace{-0.05cm}\right]^2 \\
& - \hspace{-0.1cm}\lambda_3\hspace{-0.1cm}\left[\max\left(0, \|\mathbf{W}_n\|_F^2 - P \right)\right]^2,
\end{aligned}
\end{equation}where $\lambda_\iota \gg 0$, $\iota \in \{1,2,3\}$, denotes the penalty parameter to determine the penalty magnitude\footnotemark\footnotetext{According to the penalty convergence theorem in \cite{luenberger2021linear}, the penalty method can solve the original constrained problem with either fixed or adaptive penalty parameters. In this paper, we adopt a fixed penalty scheme for ease of implementation.}.
Note that it is still generally intractable to derive a closed-form for the objective function of ($\mathrm{P}1^{'}$).
As a compromise approach, we adopt a data-driven approach to asymptotically approximate the statistical expectations involved in ($\mathrm{P}1^{'}$). Then, by exploiting the powerful capability of deep neural network (DNN) in feature extraction, we can finally obtain the solution of problem ($\mathrm{P}1^{'}$).
Based on this, the developed predictive beamforming framework is illustrated in Fig. \ref{Fig:PBF_framework}, which consists of two phases, i.e., (a) penalty method-based problem transformation and (b) DL-based problem solving.
In phase (a), we transform the original optimization problem to an unconstrained optimization problem, denoted by $\max \mathbb{E}\{f(\tb{W}_n)\}$.
In phase (b), we first adopt the Monte-Carlo method to approximate $\mathbb{E}\{f(\tb{W}_n)\}$ \cite{fishman2013monte}, i.e.,
\begin{equation}\label{}
  \mathbb{E}\{f(\tb{W}_n)\}\approx\frac{1}{N_e}\sum_{i=1}^{N_e}f(\tb{W}_n^{(i)})=\frac{1}{N_e}\sum_{i=1}^{N_e}f(g_{\omega}(\tb{\Omega}_n^{\tau(i)})).
\end{equation}
Here, the approximation holds when the number of Monte-Carlo experiments $N_e$ is sufficiently large \cite{goodfellow2016deep}. Note that $g_{\omega}(\cdot)$ denotes the DNN-based mapping function from the available input $\tb{\Omega}_n^{\tau(i)}$ to the desired output $\tb{W}_n^{(i)}$, where $\omega$ is the network parameters of DNN and $i \in \{1,2,\cdots, N_e\}$ denotes the $i$-th Monte-Carlo experiment.
Based on this, we can then set the cost function for DNN training, which is given by
\begin{equation}\label{}
  J(\omega)=-\frac{1}{N_e}\sum_{i=1}^{N_e}f(g_{\omega}(\tb{\Omega}_n^{\tau(i)})).
\end{equation}
Finally, the optimized beamforming matrix can be obtained from the DNN training by updating $\omega$ to minimize the cost function which can be expressed as
\begin{equation}\label{}
  \tb{W}_n^*=g_{\omega^*}(\tb{\Omega}_n^{\tau})
\end{equation}
with $\omega^*=\arg \min\limits_\omega J(\omega)$, where $\omega^*$ is the well-trained network parameter and $\tb{W}_n^*$ is the optimized beamforming matrix for an arbitrary $\tb{\Omega}_n^{\tau}$.

%

\textbf{Remark 2}: Note that the adopted penalty approach in phase (a) of the developed framework can be applied to any optimization problem in handling non-convex constraints. Meanwhile, the adopted DNN in phase (b) of the developed framework can be realized by any kind of neural network architecture according to the requirements, such as the convolutional neural network (CNN) \cite{lecun2015deep}, the dense neural network \cite{goodfellow2016deep}, and the residual neural network \cite{liu2020deepresidual}, etc.
Thus, the proposed framework is a universal framework for solving constrained optimization problems in a data-driven approach.

\begin{figure*}[t]
  \centering
  \includegraphics[width=0.88\linewidth]{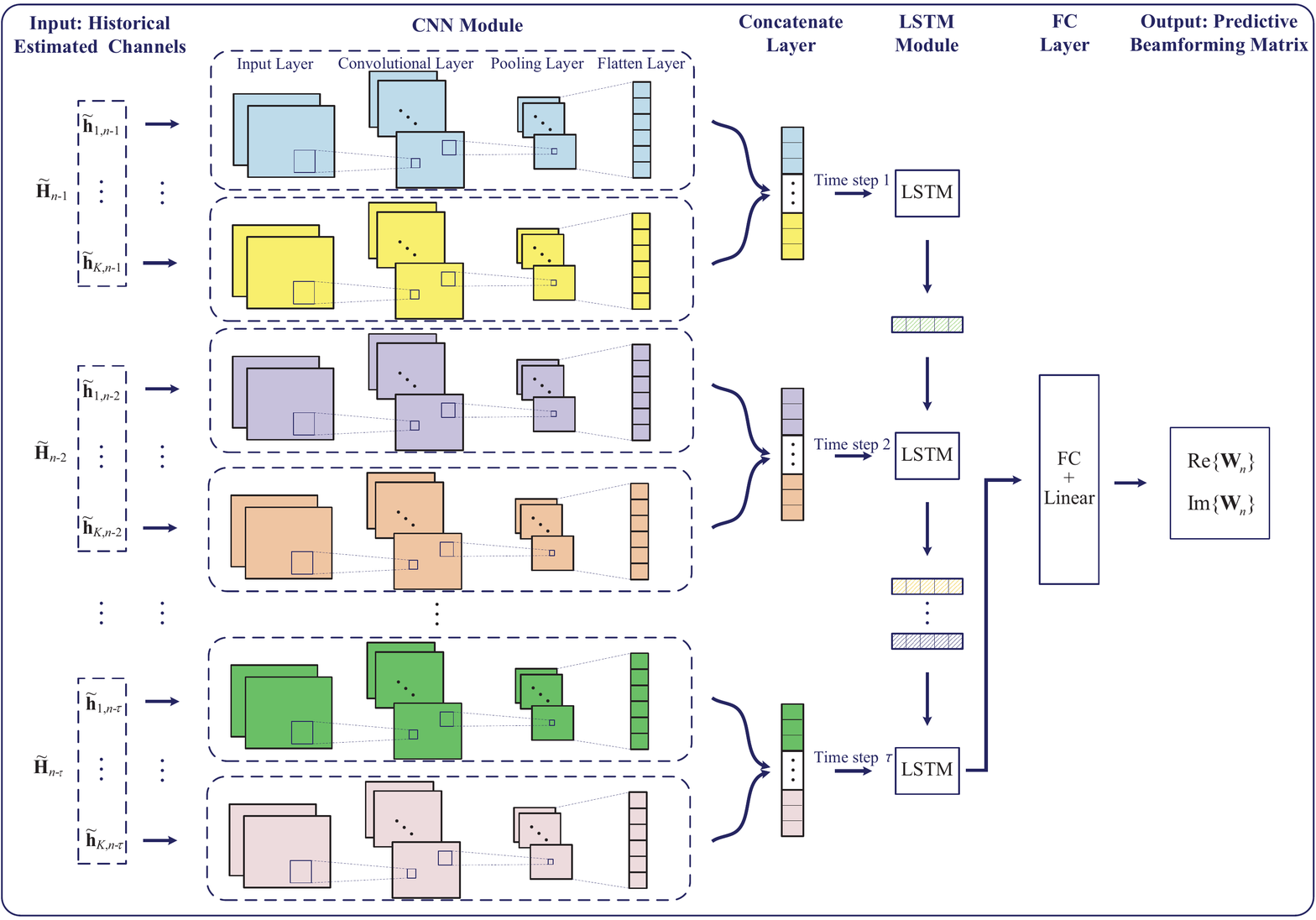}
  \caption{The developed HCL-Net architecture for the predictive beamforming in the considered V2I system.}\label{Fig:HCCL_structure}
\end{figure*}

\subsection{Proposed Historical Channels-based Convolutional LSTM Network (HCL-Net)}
As a realization of the developed framework, an HCL-Net is designed to optimize the predictive beamforming matrix, which consists of $K$ CNN modules, one concatenate layer, one LSTM module, and one fully-connected (FC) layer, as shown in Fig. \ref{Fig:HCCL_structure}.
To fully exploit the temporal dependencies\footnotemark\footnotetext{The temporal dependency here refers to the temporal correlations among historical channel vectors.} for predictive beamforming design, a classical convolutional LSTM structure \cite{goodfellow2016deep} is adopted in HCL-Net, i.e., a CNN module is first applied to extract the spatial features of the involved channel matrices to facilitate the subsequent LSTM. Then, an LSTM module is adopted to exploit the temporal dependencies of historical channels from the extracted features for predictive beamforming design.
Note that although other neural networks, such as multilayer perceptron, CNN, also have powerful capability in feature extraction, their advantages mainly lie in extracting spatial features while ignoring temporal features, which may lead to limited system performance.
In contrast to them, our proposed neural network is equipped with CNN units and memory units to exploit both spatial and temporal features to further improve the predictive beamforming performance.
The network hyperparameters are presented in Table I and the details will be introduced in the following.

\begin{table*}[t]
\normalsize
\caption{Hyperparameters of the proposed HCL-Net}\label{Tab:Hyperparameters HCL-Net}
\centering
\small
\renewcommand{\arraystretch}{1.25}
\begin{tabular}{c c c}
  \hline
   \multicolumn{3}{l}{\textbf{Input}: $\tilde{\mathbf{\Omega}}_{n}^{\tau}$ with the size of $\tau \times K \times N_t \times 2$}  \\
  \hline
  \hspace{0.1cm} \textbf{Layers/Modules} & \textbf{Parameters} &  \hspace{0.3cm} \textbf{Values}   \\
  \hspace{0.1cm} CNN module - Convolutional layer& Size of filters & \hspace{0.3cm}  $ 4 \times 3 \times 3 \times 2$   \\
  \hspace{0.1cm} CNN module - Pooling layer & Size of filters & \hspace{0.3cm}  $ 3 \times 3 $   \\
  \hspace{0.1cm} CNN module - Flatten layer & Output shape & \hspace{0.3cm}  $ 32 \times 1 $   \\
  \hspace{0.1cm} Concatenate layer & Output shape & \hspace{0.3cm}  $ 96 \times 1 $   \\
  \hspace{0.1cm} LSTM module & Size of output  & \hspace{0.3cm}  $ 64 \times 1 $   \\
  \hspace{0.1cm} FC layer & Activation function & \hspace{0.3cm} Linear \\
  \hline
   \multicolumn{3}{l}{\textbf{Output}: $[ \mathrm{Re}\{\mathbf{W}_n\}, \mathrm{Im}\{\mathbf{W}_n\} ]$}  \\
  \hline
\end{tabular}
\end{table*}

\subsubsection{\underline{Input Layer}}
To exploit the features of real-part and imaginary-part of input independently, we divide the complex-valued input into two real-valued parts, i.e.,
\begin{equation}\label{HCL_input}
  \tilde{\tb{\Omega}}_n^{\tau}= \mathcal{M}([ \mathrm{Re}\{\mathbf{\Omega}_n^{\tau}\}, \mathrm{Im}\{\mathbf{\Omega}_n^{\tau}\} ]),
\end{equation}
where $\tilde{\tb{\Omega}}_n^{\tau}$ denotes the network input and $\mathcal{M}(\cdot)\!:\mathbb{R}^{N_t \times 2\tau K}\mapsto\mathbb{R}^{\tau \times K \times N_t \times 2}$ represents the mapping function.
\subsubsection{\underline{CNN Module}}
In HCL-Net, we adopt $K$ CNN modules to independently extract the spatial features from the $K$ channel inputs at each time slot.
Specifically, each CNN module shares an identical network structure, which consists of one input layer, one convolutional layer, one pooling layer, and one flatten layer.
For the convolutional layer, we adopt 4 filters with each filter size of $3\times 3 \times 2$ to generate feature maps and a rectified linear unit (ReLU) is added after each convolution operation. Then, a maximum pooling layer with $3 \times 3$ filter size is connected to reduce the feature size. Finally, a flatten layer is adopted to generate a appropriate shape for the subsequent procedure.
\subsubsection{\underline{Concatenate Layer}}
The concatenate layer is adopted to concatenate all the extracted spatial channel features from different CNN modules. By doing this, the output of concatenate layer contains the channel features of all the $K$ vehicles at each time slot, which are fit for the subsequent temporal channel feature extraction by the LSTM module.
\subsubsection{\underline{LSTM Module}}
In the LSTM module, one LSTM unit is recurrently adopted to handle the input for the past $\tau$ time slots. Assume that there are $\tau$ time steps, the output of LSTM in time step $l-1$, $l \in \{2,3,\cdots,\tau\}$, is the input of LSTM for time step $l$. In particular, the output of LSTM in time step $\tau$ is directly adopted as the output of the entire LSTM module since it exploits the temporal dependency from all the past $\tau$ steps of historical channels.

\subsubsection{\underline{FC Layer}}
To fully exploit the extracted features from previous layers, a FC layer with a linear activation function is connected with the LSTM module to generate the desired output.

Finally, the HCL-Net can be expressed as
\begin{equation}\label{HCL_output}
  h_{\varsigma}(\tilde{\tb{\Omega}}_n^{\tau}) = [ \mathrm{Re}\{\mathbf{W}_n\}, \mathrm{Im}\{\mathbf{W}_n\} ],
\end{equation}
where $h_{\varsigma}(\cdot)$ is the non-linear function expression of HCL-Net with network parameter $\varsigma$. In this case, the optimized beamforming matrix can be written as
\begin{equation}\label{W_n_net}
  \mathbf{W}_n = \mathcal{F}(h_{\varsigma}(\tilde{\tb{\Omega}}_n^{\tau})) \in \mathbb{C}^{N_t\times K},
\end{equation}
where $\mathcal{F}(\cdot):\mathbb{R}^{N_t\times 2K}\mapsto\mathbb{C}^{N_t\times K}$ is the mapping function to generate a complex-valued matrix.

The rationale of the proposed HCL-Net can be summarized into two points:
(a) Inspired by its powerful capability in extracting both spatial and temporal features from the input, a convolutional LSTM structure is adopted in HCL-Net to further improve the learning performance; (b) We adopt $K$ four-layer CNN modules and one LSTM module in the developed HCL-Net to balance the tradeoff between the learning performance and the neural network complexity.

\textbf{Remark 3}: It should be highlighted that due to the powerful scalability of DNN, the proposed HCL-Net architecture is scalable and it can be easily extended via altering the neural network input and output sizes based on the number of vehicles, the number of transmit/receive antennas at the RSU, i.e., the proposed HCL-Net is suitable for different system deployments, which will be verified via the simulation results in Section \Rmnum{5}.

\subsection{HCL-Net-based Predictive Beamforming Algorithm}
In this section, we adopt the designed HCL-Net as the core DNN in Fig. \ref{Fig:PBF_framework} and propose an HCL-Net-based predictive beamforming algorithm, which consists of offline training and online optimization. The details of the algorithm will be introduced in the following.

\subsubsection{\underline{Offline Training}}
Given an unlabeled training set $\mathcal{X} = \left\{(\tilde{\mathbf{\Omega}}_k^{\tau(1)},\bar{\mathbf{H}}_{n}^ {(1)} ), (\tilde{\mathbf{\Omega}}_k^{\tau(2)},\bar{\mathbf{H}}_{n}^ {(2)} ), \cdots, \right.  $ $\left. ( \tilde{\mathbf{\Omega}}_k^{\tau(N_e)},\bar{\mathbf{H}}_{n}^{ (N_e)} ) \right\}$, where $N_e$ is the number of training examples, $(\tilde{\mathbf{\Omega}}_k^{\tau(i)},\bar{\mathbf{H}}_{n}^ {(i)} )$ denotes the $i$-th, $i \in \{1,2,\cdots,N_e\}$, training example in $\mathcal{X}$.
According to the developed framework in Fig. \ref{Fig:PBF_framework}, we can then define the cost function as
\begin{equation}\label{cost_function}
\begin{aligned}
  &J_{\mathrm{HCL-Net}}(\varsigma)\\
 & =  -\frac{1}{N_e}\sum_{i=1}^{N_e}
\sum_{k = 1}^K\mathrm{log}_2\left( 1 + \frac{\left| (\mathbf{h}_{k,n}^{(i)})^H\mathbf{w}_{k,n}^{(i)}(\varsigma)\right|^2 }{ {\sum_{k^{'} \neq k}^K \left| \mathbf{h}_{k,n}^H\mathbf{w}_{k^{'},n}\right|^2 + \sigma_{k}^2}}\right) \\
& + \lambda_1 \left[\max\left(0,\frac{1}{N_eK}\sum_{i=1}^{N_e} \sum_{k=1}^K\mathrm{CRLB}(\theta_{k,n}^{(i)},\tb{w}_{k,n}^{(i)}(\varsigma)) - \gamma_{\theta} \right)\right]^2 \\
& + \lambda_2\left[\max\left(0, \frac{1}{N_eK}\sum_{i=1}^{N_e} \sum_{k=1}^K\mathrm{CRLB}(d_{k,n}^{(i)},\tb{w}_{k,n}^{(i)}(\varsigma)) - \gamma_d \right)\right]^2 \\
& + \lambda_3 \frac{1}{N_e}\sum_{i=1}^{N_e} \left[\max\left(0, \|\mathbf{W}_{n}^{(i)}(\varsigma)\|_F^2 - P \right)\right]^2.
\end{aligned}
\end{equation}Here, the corresponding $\mathrm{CRLB}(\cdot,\cdot)$ were defined in (\ref{CRLB_theta}) and (\ref{CRLB_d}), respectively, and $\mathbf{w}_{k,n}^{(i)}(\varsigma)$ is the $k$-th column of $\mathbf{W}_{n}^{(i)}(\varsigma)=\mathcal{F}(h_{\varsigma}(\tilde{\tb{\Omega}}_n^{\tau(i)}))$, as defined in (\ref{W_n_net}). For the maximum operators, i.e., $\max(\cdot,\cdot)$ in (\ref{cost_function}), a rectified linear unit (ReLU) function \cite{goodfellow2016deep}, i.e., $f_{\mathrm{R}}(x) = \max(0,x)$ can be adopted to replace them to facilitate the network training.
Based on (\ref{cost_function}), we can then adopt the back propagation algorithm (BPA) to update the network parameters $\varsigma$ progressively to minimize the cost function value.
After the training process, a well-trained HCL-Net can be expressed as
\begin{equation}\label{well_trained_HCL}
  h_{\varsigma^*}(\tilde{\tb{\Omega}}_n^{\tau}) = [ \mathrm{Re}\{\mathbf{W}_n^*\}, \mathrm{Im}\{\mathbf{W}_n^*\} ], \forall n,
\end{equation}
where $h_{\varsigma^*}(\cdot)$ denotes the well-trained HCL-Net with the well-trained network parameters $\varsigma^*$, and $\mathbf{W}_n^*$ denotes the optimized predictive beamforming matrix based on the HCL-Net.

\begin{table}[t]
\small
\centering
\begin{tabular}{l}
\toprule[1.8pt] \vspace{-0.3 cm}\\
\hspace{-0.1cm} \textbf{Algorithm 1} {HCL-Net-based Predictive Beamforming Algorithm} \vspace{0.2 cm} \\
\toprule[1.8pt] \vspace{-0.3 cm}\\
\textbf{Initialization:} $i_t = 0$, and $I_t = N_{\max}$ \\
\hspace{2.2cm}$\varsigma$ with random weights \\
\hspace{2.2cm}a training set $\mathcal{X}$ \\
\textbf{Unsupervised Offline Training:} \\
1:\hspace{0.75cm}\textbf{Input:} Training set $\mathcal{X}$\\
2:\hspace{1.1cm}\textbf{while} $i_t \leq I_t $ \textbf{do} \\
3:\hspace{1.6cm}Update $\varsigma$ by BPA to minimize $J_{\mathrm{HCL-Net}}(\varsigma)$ in (\ref{cost_function}) \\
\hspace{1.8cm} $i_t = i_t + 1$  \\
4:\hspace{1.1cm}\textbf{end while} \\
5:\hspace{0.75cm}\textbf{Output}:  Well-trained ${h}_{\varsigma^*}( \cdot ) $ as defined in (\ref{well_trained_HCL})\\
\textbf{Online Beamforming Design:} \\
6:\hspace{0.75cm}\textbf{Input:} Test data  $\tilde{\tb{\Omega}}_m^{\tau}= \mathcal{M}([ \mathrm{Re}\{\mathbf{\Omega}_m^{\tau}\}, \mathrm{Im}\{\mathbf{\Omega}_m^{\tau}\} ])$ \\
7:\hspace{1.1cm}\textbf{do} Predictive Beamforming  using \\
\hspace{1.7cm} well-trained HCL-Net ${h}_{\varsigma^*}( \cdot ) $ \\
8:\hspace{0.75cm}\textbf{Output:} $\mathbf{W}_m^* = \mathcal{F}(h_{\varsigma^*}(\tilde{\tb{\Omega}}_m^{\tau}))$ \vspace{0.2cm}\\
\bottomrule[1.8pt]
\end{tabular}
\end{table}

\subsubsection{\underline{Online Optimization}}
Given a test example $\tb{\Omega}_m^{\tau}$, $m \neq n$, we can obtain the network input, i.e., $\tilde{\tb{\Omega}}_m^{\tau}= \mathcal{M}([ \mathrm{Re}\{\mathbf{\Omega}_m^{\tau}\}, \mathrm{Im}\{\mathbf{\Omega}_m^{\tau}\} ])$. Then, we send $\tilde{\tb{\Omega}}_m^{\tau}$ into the well-trained HCL-Net to obtain the optimized predictive beamforming matrix:
\begin{equation}\label{W*_n_net_test}
  \mathbf{W}_m^* = \mathcal{F}(h_{\varsigma^*}(\tilde{\tb{\Omega}}_m^{\tau})).
\end{equation}

\subsubsection{ \underline{Algorithm Steps}}
Based on the above discussion, we then summarize the proposed algorithm steps in \textbf{Algorithm 1}, where $i_t$ is the iteration index and $I_t = N_{\max}$ is the maximum iteration number.

\subsection{Complexity Analysis}
The complexity of the proposed algorithm consists of two parts: offline training and online beamforming. In fact, both the offline and online processes require the calculation of the proposed neural network, which mainly consists of two parts: the calculations of the CNN module and the LSTM module.
According to \cite{he2015convolutional}, the computation of a CNN module at each time step is $\mathcal{O}\left( \sum_{l = 1}^{N_L}n_{l-1}  s_l^2  n_l a_l b_l\right)$.
Here, $\mathcal{O}(\cdot)$ represents the order of the computational complexity, $N_L$ represents the number of convolutional layers, and $n_0$ is the input dimension.
In addition, subscript $l$, $l \in \{1,2,\cdots, N_l\}$, denotes the index of the convolutional layer, and $n_l$ and $s_l$ are the number of neural network channels and the spatial size of the filter, respectively. Moreover, $a_l$ and $b_l$ are the length and the width of the output feature map.
Note that we adopt $K$ CNN modules to independently handle the channel inputs from the $K$ vehicles at each time step and there are $\tau$ time steps in the proposed neural network.
In this case, the complexity of the calculation of CNN modules is $\mathcal{O}\left( \tau K\sum_{l = 1}^{N_L}n_{l-1}  s_l^2  n_l a_l b_l\right)$.
Concerning the complexity of the LSTM module, one LSTM unit is recurrently adopted for all the $\tau$ time steps.
According to \cite{hochreiter1997long, tsironi2017analysis}, the complexity of an LSTM unit per time step is $\mathcal{O}\left( 4(\kappa_1\kappa_2 + \kappa_2^2 + \kappa_2 )\right)$, where $\kappa_1$ and $\kappa_2$ represent the input dimension and the output dimension of the LSTM unit, respectively.
Since there are $\tau$ time steps in the proposed neural network, the complexity of the calculation of the LSTM module is $\mathcal{O}\left( 4\tau(\kappa_1\kappa_2 + \kappa_2^2 + \kappa_2 )\right)$.

Based on the above discussion, the complexity of the offline training of the proposed algorithm is
\begin{equation}\label{}
\begin{aligned}
  &C_\mathrm{offline} \\
  &= \mathcal{O}\hspace{-0.05cm}\left(\hspace{-0.05cm} I_tN_e\hspace{-0.05cm}\left(\hspace{-0.05cm}\tau K\sum\limits_{l = 1}^{N_L}n_{l-1} s_l^2  n_l a_l b_l + 4\tau(\kappa_1\kappa_2 + \kappa_2^2 + \kappa_2 )\hspace{-0.05cm}\right)\hspace{-0.05cm}\right),
\end{aligned}
\end{equation}
where $I_t$ and $N_e$ are the maximum iteration number and the number of training examples, respectively, as defined in Section IV-C.
Moreover, the complexity of the online beamforming of the proposed algorithm is
\begin{equation}\label{}
  C_\mathrm{online} = \mathcal{O}\left( \tau K\sum\limits_{l = 1}^{N_L}n_{l-1}  s_l^2  n_l a_l b_l + 4\tau(\kappa_1\kappa_2 + \kappa_2^2 + \kappa_2 )\right).
\end{equation}

\begin{figure}[t]
  \centering
  \includegraphics[width=\linewidth]{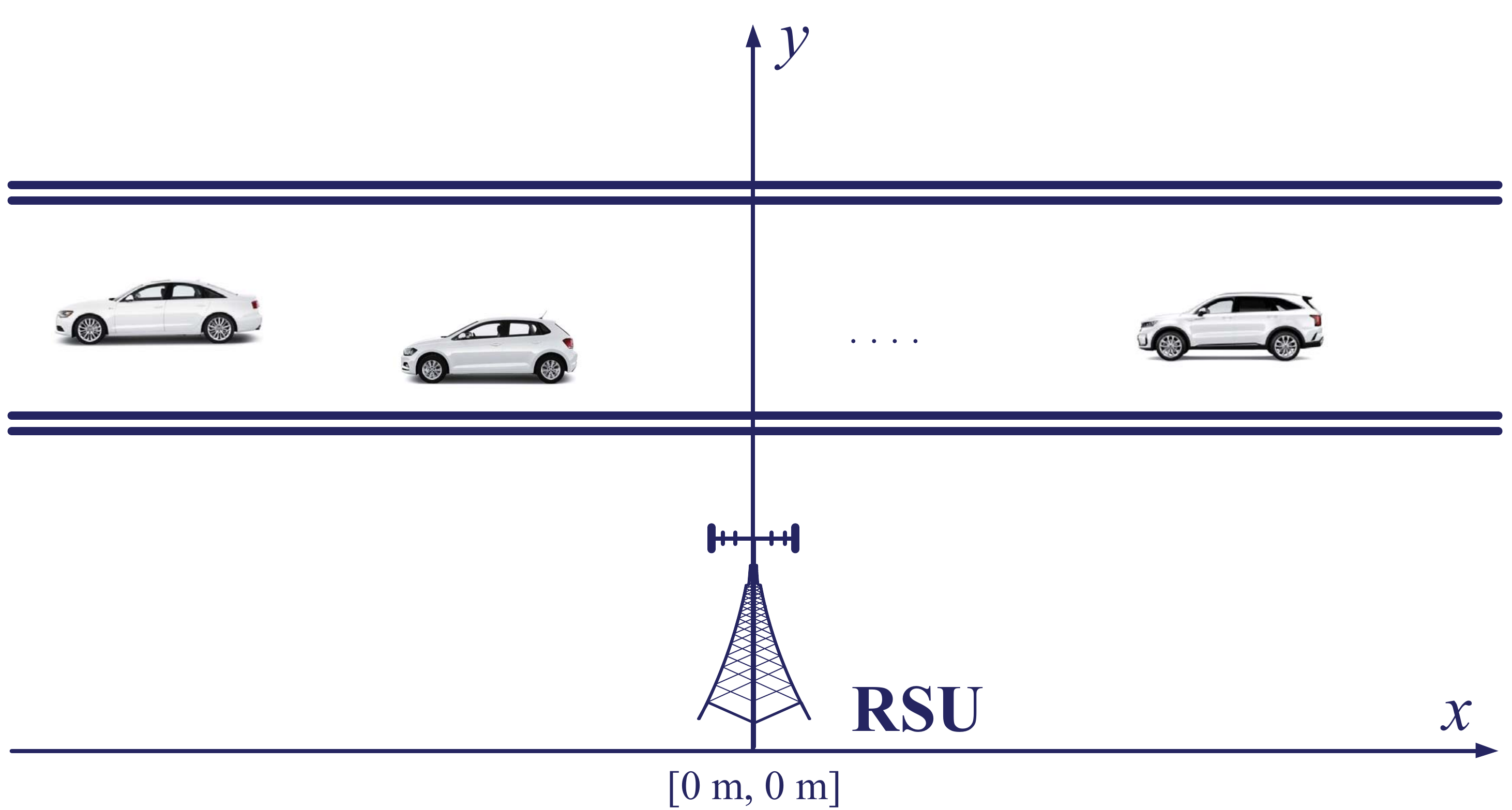}
  \caption{The considered V2I system for simulations.}\label{Fig:scenario_simulations}
\end{figure}

\section{Numerical Results}
In this section, we conduct simulations in an ISAC-assisted V2I network under mmWave setting and provide extensive numerical results to verify the effectiveness of the proposed method.
In the considered V2I network, one RSU equipped with $N_t$ transmit antennas and $N_r$ receive antennas serves $K$ single-antenna vehicles, as defined in the system model.
Unless otherwise specified, we set $N_t = N_r = 32$, $K = 3$, $\sigma_z^2 = \sigma_k^2 = -80~\mathrm{dBm}$.
Fig. \ref{Fig:scenario_simulations} adopts the two-dimensional (2D) coordinate system to illustrate the relative positions of the RSU and the $K$ vehicles, where the RSU is located at $[0~\mathrm{m}, 0~\mathrm{m}]$.
Without loss of generality, the initial positions of vehicles are set randomly, which are formulated as $[x_k,y_k] = [\tilde{x}_k + \Delta x,\tilde{y}_k + \Delta y]$. Here, $[x_k, y_k]$ denotes the coordinate of the $k$-th vehicle, $[\tilde{x}_k,\tilde{y}_k]$ is the mean initial position, and $\Delta x, \Delta y \sim \mathcal{N}(0,1)$ are random variables to characterize the uncertainty of  initial positions.
Accordingly, we set $\Delta T = 0.02~\mathrm{s}$ and the average velocity of the $k$-th vehicle is set as $v_{k,n} \sim \mathcal{U}(8~\mathrm{m/s}, 8.25~\mathrm{m/s})$, i.e., the average velocity for vehicles is around $30~\mathrm{km/h}$.
For the observation model, we set $\xi= 10$ and $\rho_{\nu} = 2.0 \times 10^{-6}$.
For the optimization problem, we set $\tau = 6 $, $\gamma_{\theta} = 0.01~\mathrm{rad}^2$, $\gamma_d = 0.01~\mathrm{m}^2$, and $\lambda_1 = \lambda_2 = \lambda_3 = 10^3$.
In addition, the historical estimated motion parameters are set with a normalized mean squared error (NMSE) of $0.01$.
The other default settings for simulations are presented in Table \ref{Tab:Simulation_settings}.


\begin{table}[t]
\normalsize
\caption{Default Settings in Simulations}\label{Tab:Simulation_settings}
\centering
\small
\renewcommand{\arraystretch}{0.6}
\begin{tabular}{c c}
  \hline
  \\
   \textbf{Parameters} &  \textbf{Default Values} \vspace{0.2cm} \\
  \hline
  \\
   Path loss exponent & $\zeta = 2.55$  \\
   \\
   Path loss \cite{niu2015survey} at $d_0 = 1~\mathrm{m}$ & $\alpha_0 = -70~\mathrm{dB}$  \\
   \\
   Fading coefficient \cite{yuan2020bayesian} & $\varrho = 10 + 10j$  \\
   \\
   Position of RSU & $[0~\mathrm{m},0~\mathrm{m}]$ \\
   \\
   Mean initial position of vehicle 1 & $[15~\mathrm{m},20~\mathrm{m}]$ \\
   \\
   Mean initial position of vehicle 2 & $[25~\mathrm{m},20~\mathrm{m}]$ \\
   \\
   Mean initial position of vehicle 3 & $[35~\mathrm{m},20~\mathrm{m}]$ \\
   \\
   Carrier frequency & $f_c = 30~\mathrm{GHz}$ \\
   \\
   Speed of signal propagation & $c = 30~\mathrm{GHz}$ \\
   \\
  Maximum CRLB threshold (angle) & $\gamma_{\theta} = 0.01~\mathrm{rad}^2$\\
  \\
  Maximum CRLB threshold (distance) & $\gamma_d = 0.01~\mathrm{m}^2$ \\
   \\
  \hline
\end{tabular}
\end{table}

To evaluate the beamforming performance of the proposed method, we adopt the following three methods as benchmarks for comparisons:
\begin{itemize}
  \item Benchmark 1 (Upper bound): It is a genie-aided scheme in which downlink multiple access interference does not exist. Besides, constraints (\ref{P1_C_theta}) and (\ref{P1_C_d}) are ignored. Also, perfect ICSI is available such that the optimal beamforming can be obtained \cite{rao2003performance} to achieve the upper bound performance of problem (P1).
  \item Benchmark 2 (Naive DL method): In this scheme, only the estimated channels at time slot $n-1$, i.e., $\tilde{\mathbf{H}}_{n-1}$ is available and this method naively regards $\tilde{\mathbf{H}}_{n-1}$ as the channel at time slot $n$. In contrast to the proposed method, this method adopts a fully-connected neural network to extract features from $\tilde{\mathbf{H}}_{n-1}$ for beamforming design at time slot $n$.
  \item Benchmark 3 (Random beamforming method): A random beamforming scheme is adopted, where the beamforming matrix is set randomly without the consideration of constraints (\ref{P1_C_theta}) and (\ref{P1_C_d}).
\end{itemize}
For the proposed method, the adopted hyperparameters for simulations are based on Table \ref{Tab:Hyperparameters HCL-Net}, the numbers of training examples and test examples are both set as $2,000$\footnotemark\footnotetext{In the simulations, since the illustrative system setting is not complicated and the adopted HCL-Net is with few parameters, only a relatively small number of training examples are required \cite{goodfellow2016deep}. Indeed, our simulation results show that a training set with a total number of $2,000$ examples is sufficient to achieve an excellent performance.}, and the number of epochs \cite{goodfellow2016deep}, i.e., the number of times that the learning algorithm works through the entire training dataset, is set as $6$.
In addition, each point in the simulation results is obtained via averaging over $2,000$ Monte Carlo realizations.
In the following, we will conduct simulations in terms of communication performance, sensing performance, and the neural network training performance, respectively.

\begin{figure}[t]
  \centering
  \includegraphics[width=3in,height=2.6in]{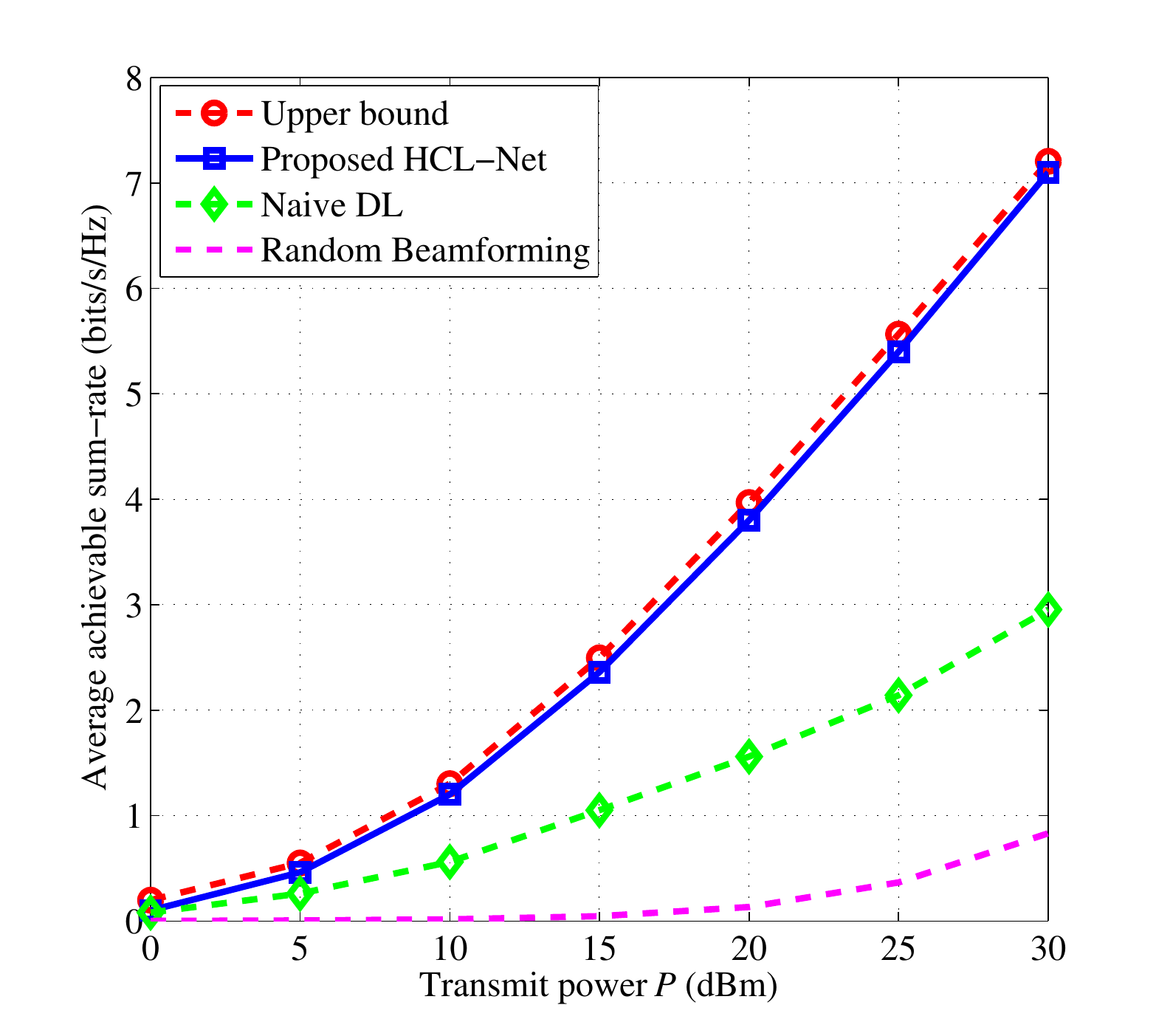}
  \caption{The average achievable sum-rate with different transmit powers under $N_t = N_r = 32$.}\label{Fig:Rate_P}
\end{figure}

\subsection{Communication Performance}
To evaluate the communication performance, we investigate the achievable sum-rates of V2I network by varying different power budgets, number of transceiver antennas, and number of vehicles, respectively.
Fig. \ref{Fig:Rate_P} presents the average achievable sum-rate versus the maximum transmit power $P$ at the RSU. It is shown that the achievable sum-rates of all the considered algorithms increase with $P$. This is because a higher power budget improves the received signal strengths at vehicles.
In particular, the random beamforming method presents a very limited performance since the random beamformer fails to align with the desired channels between the RSU and vehicles. On the other hand, although the naive DL method outperforms that of the random beamforming method, it only offers a small performance gain. The reason is that for the naive DL method, only the outdated channel information is available for optimizing the beamforming matrix, which can hardly track the high-speed vehicles in V2I networks leading to inefficient beamforming.
Different from these two methods, we can observe that the proposed method presents a significant performance improvement and its achievable sum-rate performance approaches closely to that of the upper bound exploiting perfect ICSI.
Specifically, the sum-rate of our proposed method enjoys the same slope as the upper bound method in exploiting the transmit power for performance improvement. In particular, a $10~\mathrm{dB}$ of performance gain is achieved when the data rate is $3~\mathrm{bits/s/Hz}$ compared with the naive DL method.
This is expected since our proposed method can intelligently predict the beamforming matrix via exploiting features from the historical estimated channels to further improve the sum-rate performance.

\begin{figure}[t]
  \centering
  \includegraphics[width=3in,height=2.6in]{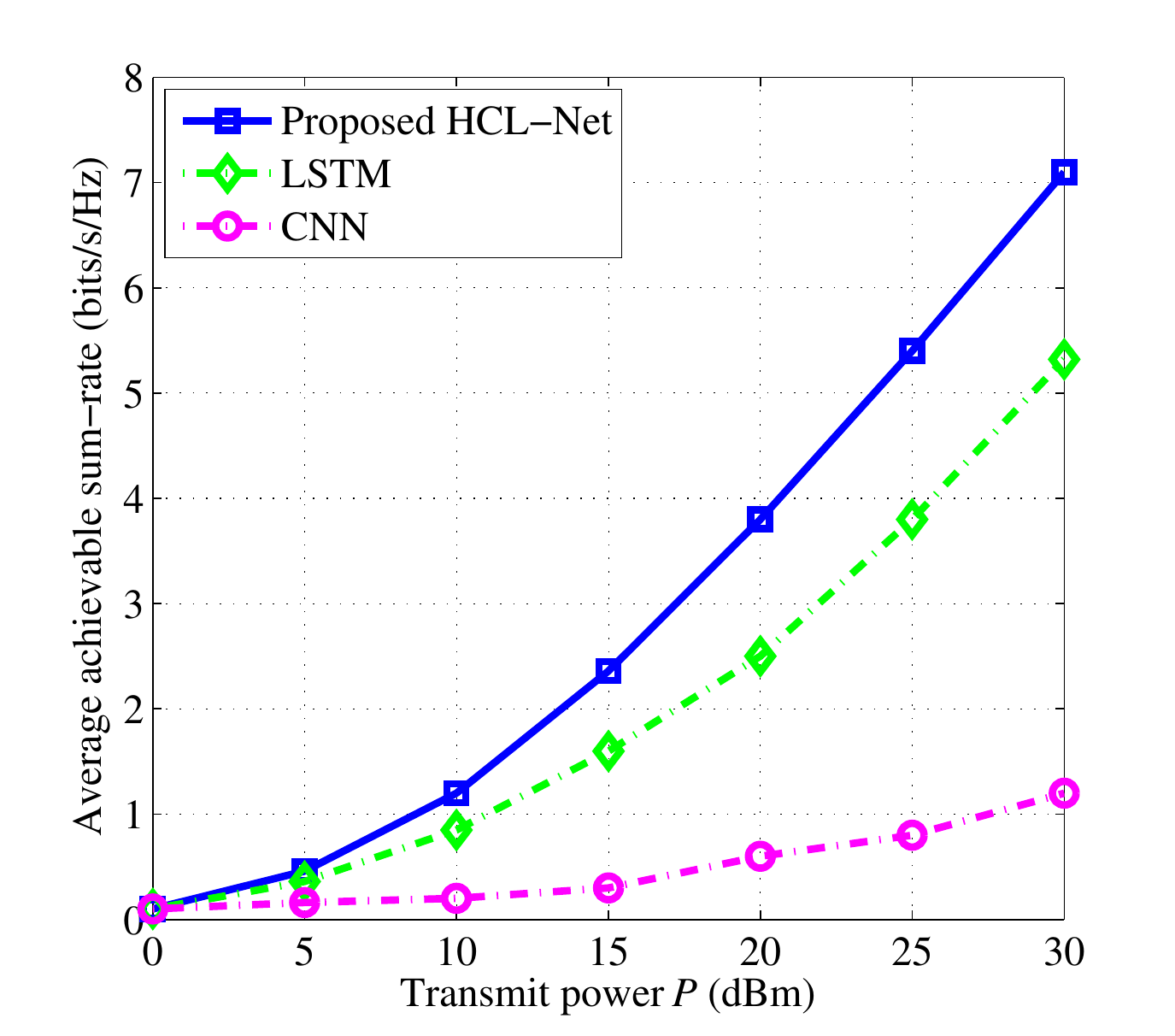}
  \caption{The average achievable sum-rate with different transmit powers under $N_t = N_r = 32$.}\label{Fig:Reviewer1_rate_power_comparison}
\end{figure}

\begin{figure}[t]
  \centering
  \includegraphics[width=3in,height=2.6in]{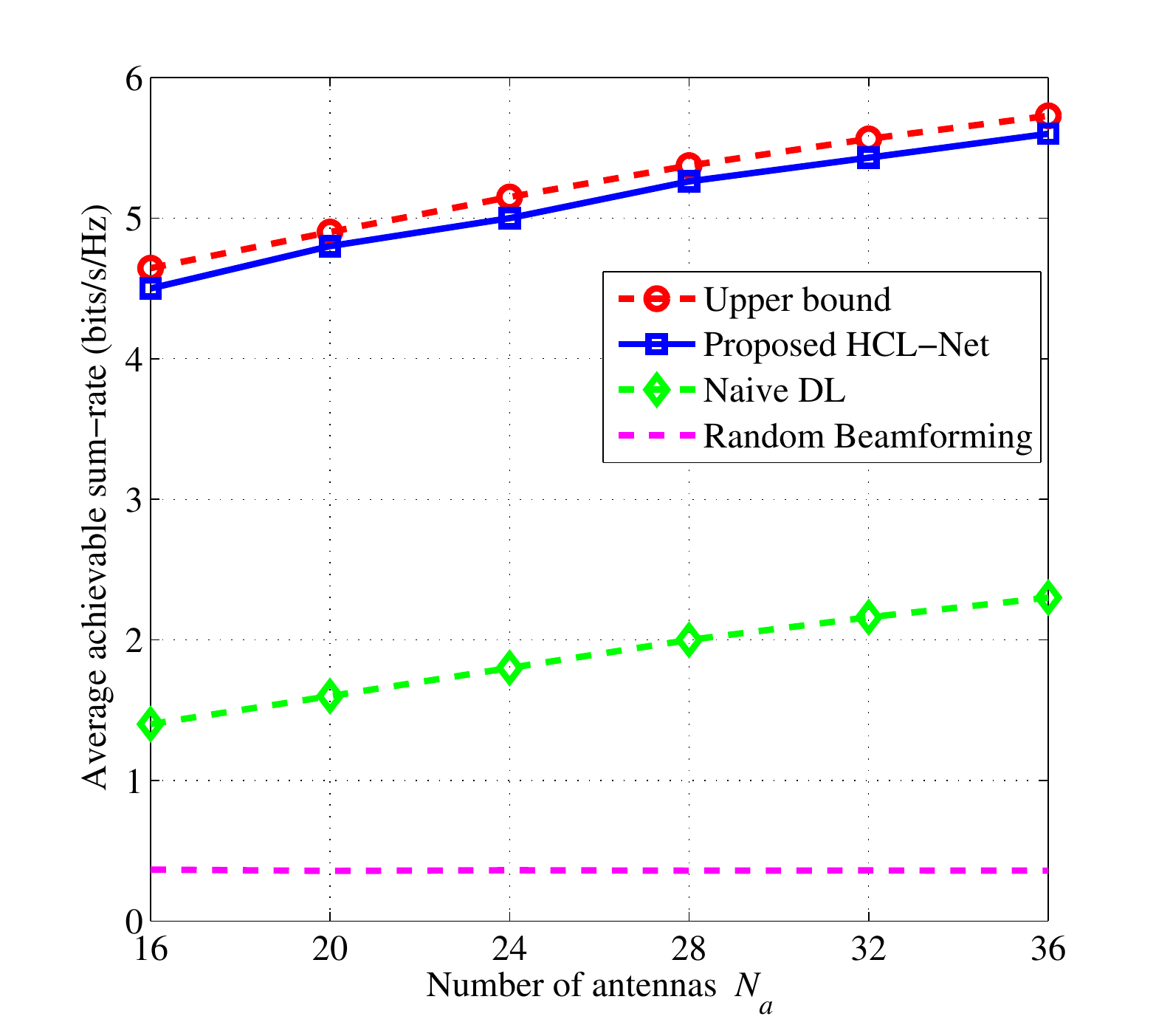}
  \caption{The average achievable sum-rate with different number of antennas under $P = 25~\mathrm{dBm}$ and $N_t = N_r = N_a$.}\label{Fig:Rate_N}
\end{figure}

To study the impacts of different neural network structure on the sum-rate performance, we compare our proposed HCL-Net with the classical LSTM structure and CNN structure.
As shown in Fig. \ref{Fig:Reviewer1_rate_power_comparison}, the proposed HCL-Net method achieves the best sum-rate performance and outperforms the other two neural networks significantly.
The reason is that both LSTM and CNN can only exploit either temporal feature or spatial feature, while the proposed HCL-Net can leverage the CNN and LSTM modules to simultaneously exploit both the spatial feature and the temporal feature of communication channels to further improve the sum-rate performance. Therefore, although the proposed HCL-Net has a larger computational complexity compared with LSTM and CNN, the former can achieve an excellent sum-rate performance.

\begin{figure}[t]
\centering
\subfigure[]{
\includegraphics[width=3in,height=2.6in]{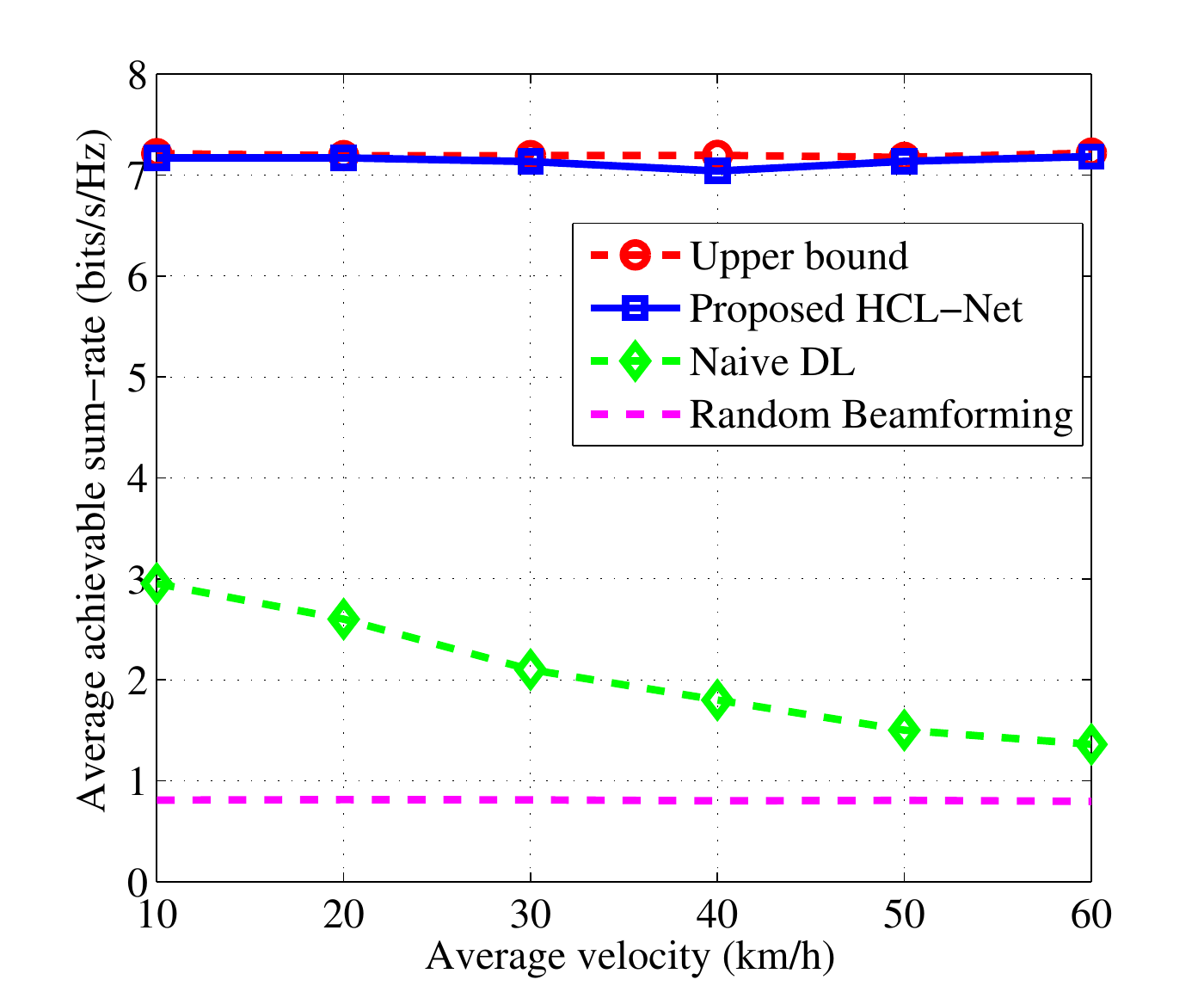}
\label{Fig:rate_speed}}
\\
\subfigure[]{
\includegraphics[width=3in,height=2.6in]{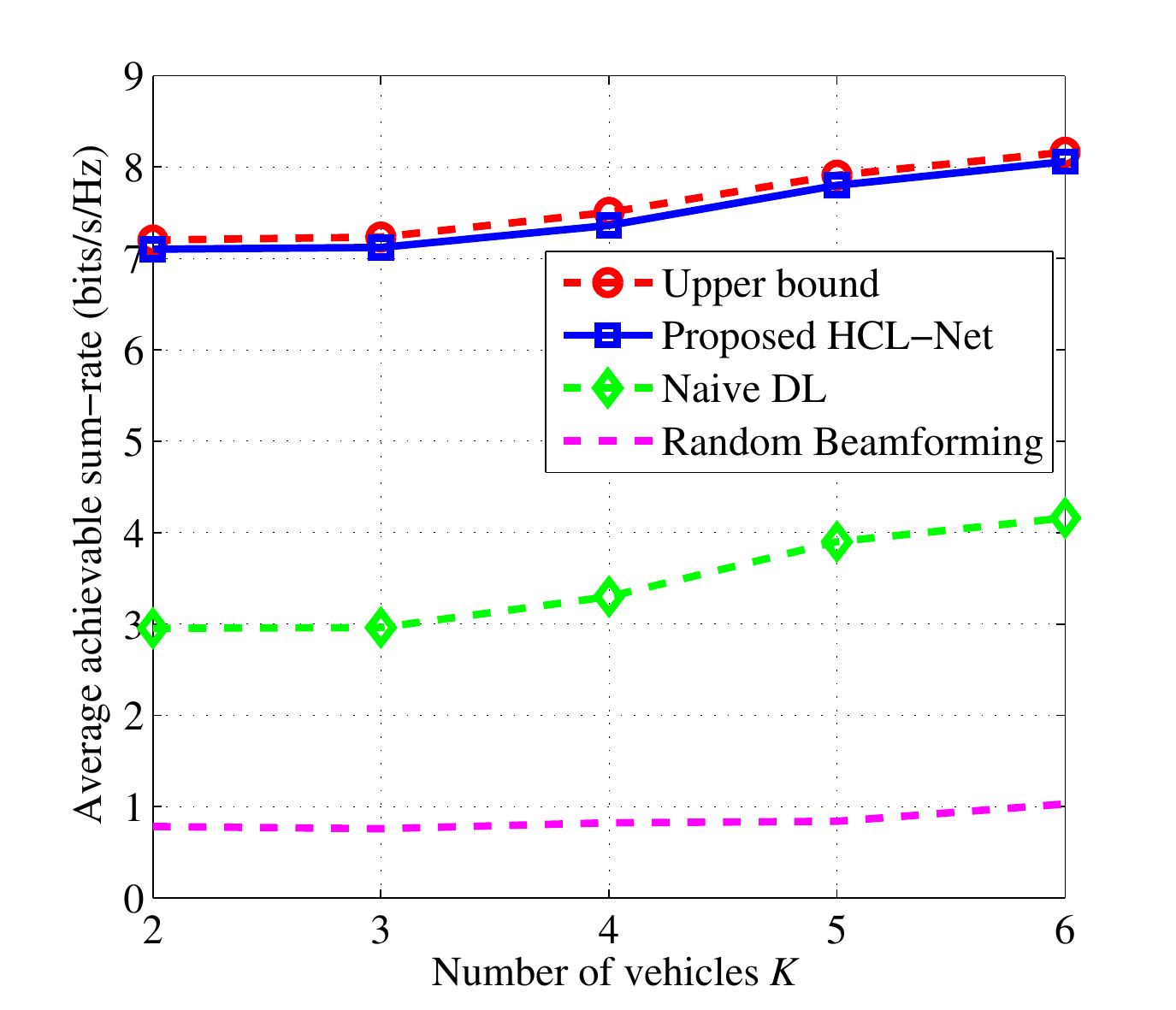}
\label{Fig:rate_vehicles}}
\\
\caption{The effect of the vehicles' parameters on the ISAC system performance under $N_t = N_r = 32$ and $P = 30~\mathrm{dBm}$. (a) The average achievable sum-rate versus the average velocity of vehicles; (b) The average achievable sum-rate versus the number of vehicles.}
\end{figure}

On the other hand, the number of transmit/receive antennas at the RSU also has some important impacts on the sum-rate performance.
For ease of study, we set $N_t = N_r = N_a$, where $N_a$ denotes the number of transmit/receive antennas at the RSU.
Then, we fix $P = 25~\mathrm{dBm}$ and present the simulation results of the average achievable sum-rate under different number of antennas in Fig. \ref{Fig:Rate_N}.
It can be seen that when the number of antennas equipped at the RSU increases, the random beamforming method only achieves a quite small sum-rate gain, while all the other methods can achieve remarkable performance gain.
The reason is that the random beamforming policy can hardly exploit the degrees-of-freedom in the spatial domain for focusing the information carrying beams on the desired directions for interference management.
Similar to the results in Fig. \ref{Fig:Rate_P}, our proposed method outperforms the naive DL method and the random beamforming method significantly while achieving a considerable performance of the upper bound method.
For example, when $N_t = N_r = 28$, the average achievable sum-rate of our proposed method is around $5~\mathrm{bits/s/Hz}$, which is $1.5$ times higher than that of the naive DL method.
In fact, our proposed HCL-Net method can exploit the DNN's powerful scalability, thus presenting satisfactory performance under different number of antennas.

\begin{figure}[t]
\centering
\subfigure[]{
\includegraphics[width=3in,height=2.6in]{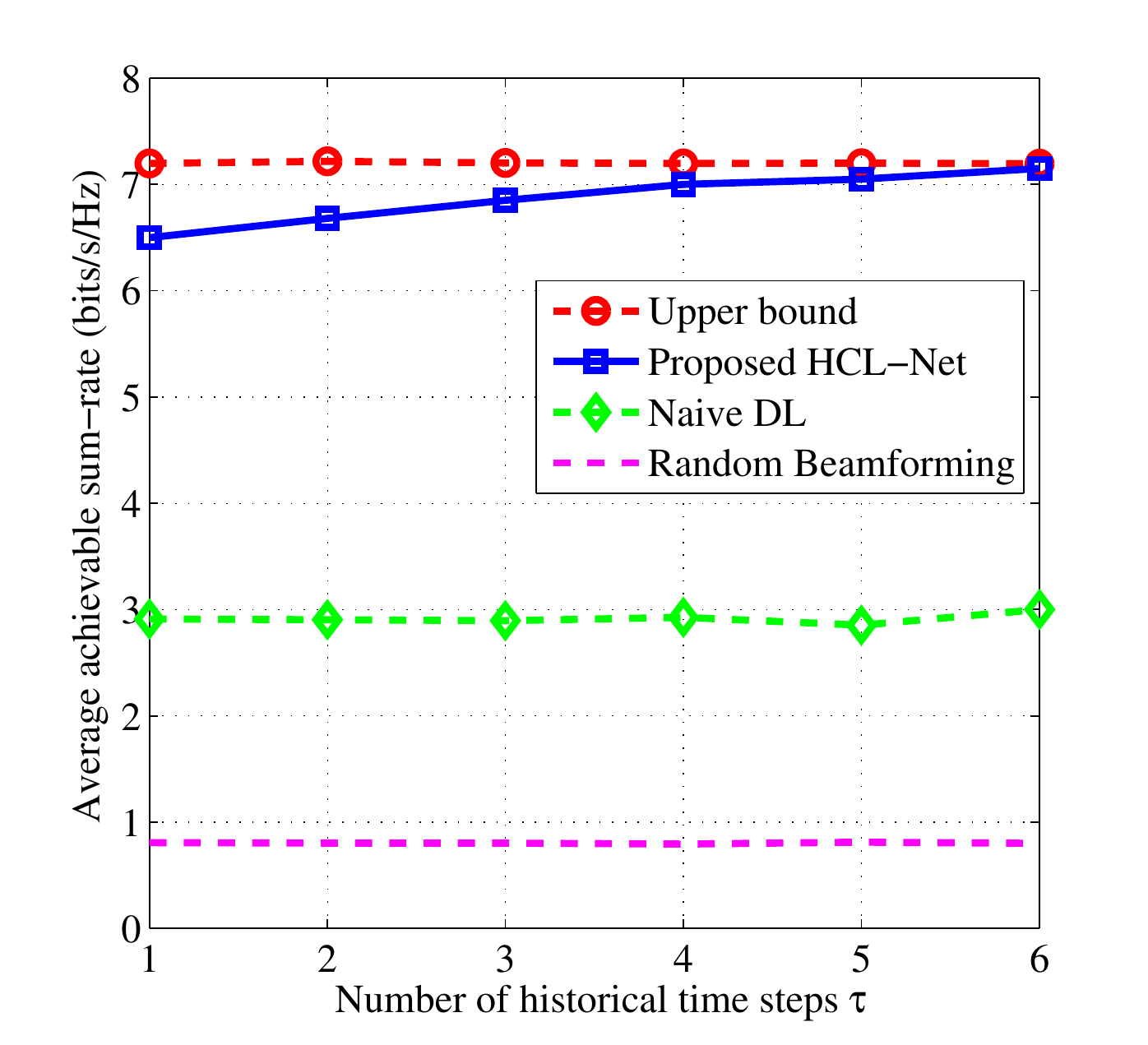}
\label{Fig:rate_time_steps}}
\\
\subfigure[]{
\includegraphics[width=3in,height=2.6in]{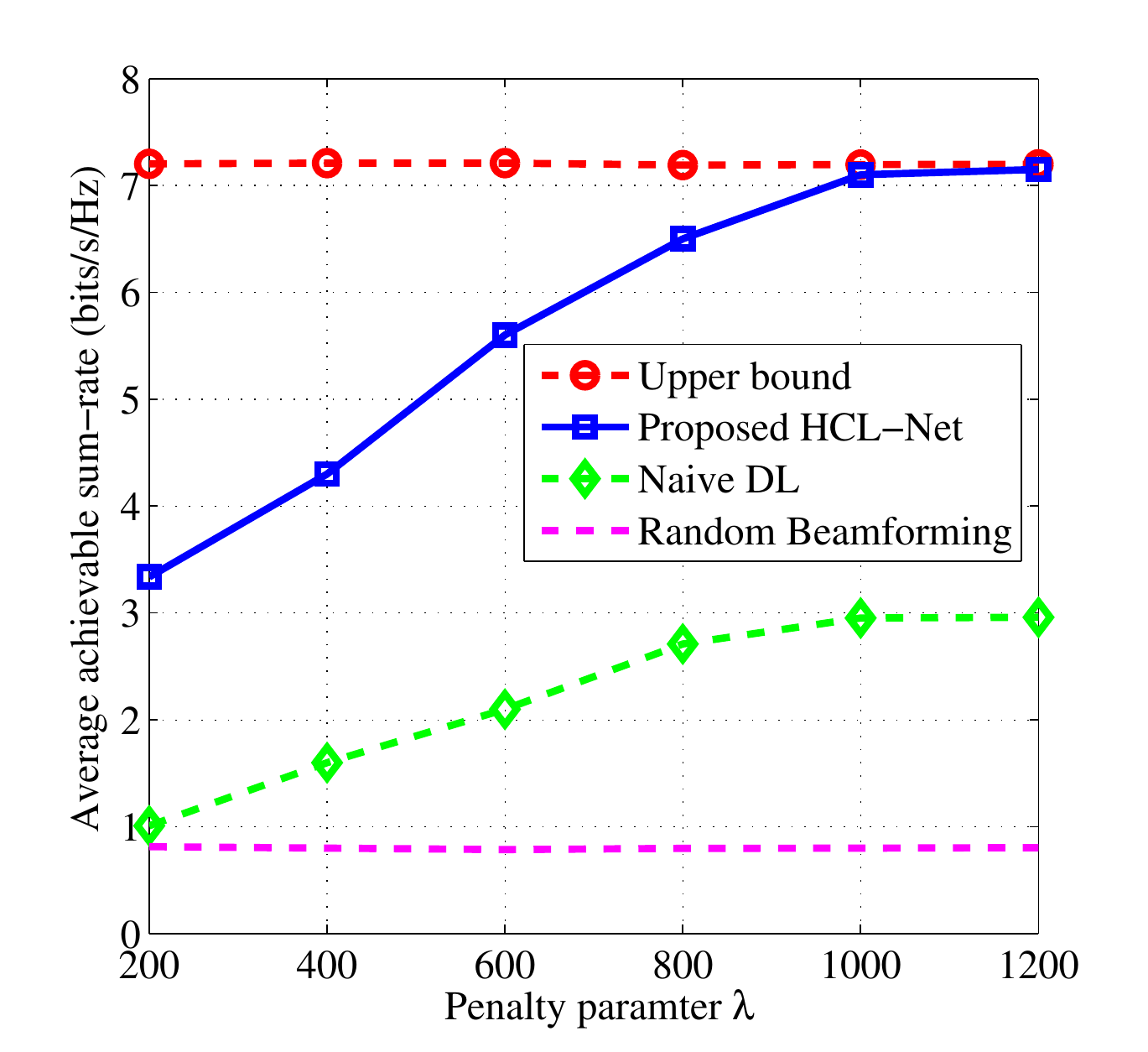}
\label{Fig:rate_lambda}}
\caption{The effect of the algorithm's parameters on the ISAC system performance under $N_t = N_r = 32$ and $P = 30~\mathrm{dBm}$. (a) The average achievable sum-rate versus the number of historical time steps; (b) The average achievable sum-rate versus the penalty parameters with $\lambda_1 = \lambda_2 = \lambda_3=\lambda$.}
\end{figure}

In the following, we will investigate the effect of the vehicles' parameters on the communication performance.
Fig. \ref{Fig:rate_speed} presents the curves of the achievable sum-rate versus the average velocity, where we vary the velocity from $10~\mathrm{km/h}$ to $60~\mathrm{km/h}$ to obtain the test results.
It can be observed that for the upper bound method and the random beamforming method, the achievable sum-rate almost keeps the same value under different velocity values since they do not need to explore the temporal dependency of channels for beamforming design.
In addition, the sum-rate of the naive method decreases with the increase of the vehicles' average velocity. The reason is that the naive method only exploits the current channel for the beamforming design in the future time slot, while the increase of the velocity would make the future channel differs significantly from the current channel.
In contrast to the naive method, the performance of our proposed method can still approach that of the upper bound under different velocity values since it can leverage deep learning technology to implicitly and accurately predict the channel in the next time slot for predictive beamforming design.
The curves of the achievable sum-rate under different number of vehicles are shown in Fig. \ref{Fig:rate_vehicles}, where we vary the size of the proposed HCL-Net according to the number of vehicles $K$ to test its performance.
The results show that the achievable sum-rate increases with the number of vehicles since the increased number of vehicles can introduce certain multi-user diversity gain.
In addition, the sum-rate performance of the proposed method can still approach that of the upper bound method while outperforming the other two methods significantly.
The reason is that the proposed algorithm can exploit the powerful scalability of DNN to efficiently exploit the multi-user diversity gain to facilitate the beamforming design.

Besides, we study the effect of the algorithm's parameters on the system performance.
The simulation results of the achievable sum-rate versus number of historical time steps $\tau$ is presented in Fig. \ref{Fig:rate_time_steps}.
It can be observed that for the upper bound method, the naive DL method, and the random beamforming method, their sum-rate performance almost keep constant when $\tau$ changes.
This is as expected since these three methods do not exploit the temporal dependency for beamforming design.
In contrast, the performance of our proposed method increases with $\tau$ since the developed HCL-Net can fully exploit the temporal dependency from the historical channels for beamforming design to further improve the system performance.
On the other hand, to study the effect of the penalty parameters on the system performance, we let $\lambda_1 = \lambda_2 = \lambda_3=\lambda$ and vary the value of $\lambda$ to obtain the simulation results, as shown in Fig. \ref{Fig:rate_lambda}.
It can be observed that the performance of the upper bound method and the random beamforming method does not change with $\lambda$ since they do not adopt the penalty method.
In addition, we can find that for the proposed method and the naive DL method, the achievable sum-rate increases with $\lambda$.
This is as expected since if $\lambda$ is small, it is easy to yield large constraint violations, which results in infeasible solutions and eventually leads to a low average achievable sum-rate.
Therefore, a larger valued $\lambda$ contributes to the improvement of the achievable sum-rate.
In particular, the achievable sum-rate converges to the optimal upper bound after $\lambda\geq 1000$ in all the considered cases, which verified that the setting of $\lambda_1 = \lambda_2 = \lambda_3 = 1000$ in our simulations is reasonable.

\begin{figure}[t]
  \centering
  \includegraphics[width=3in,height=2.6in]{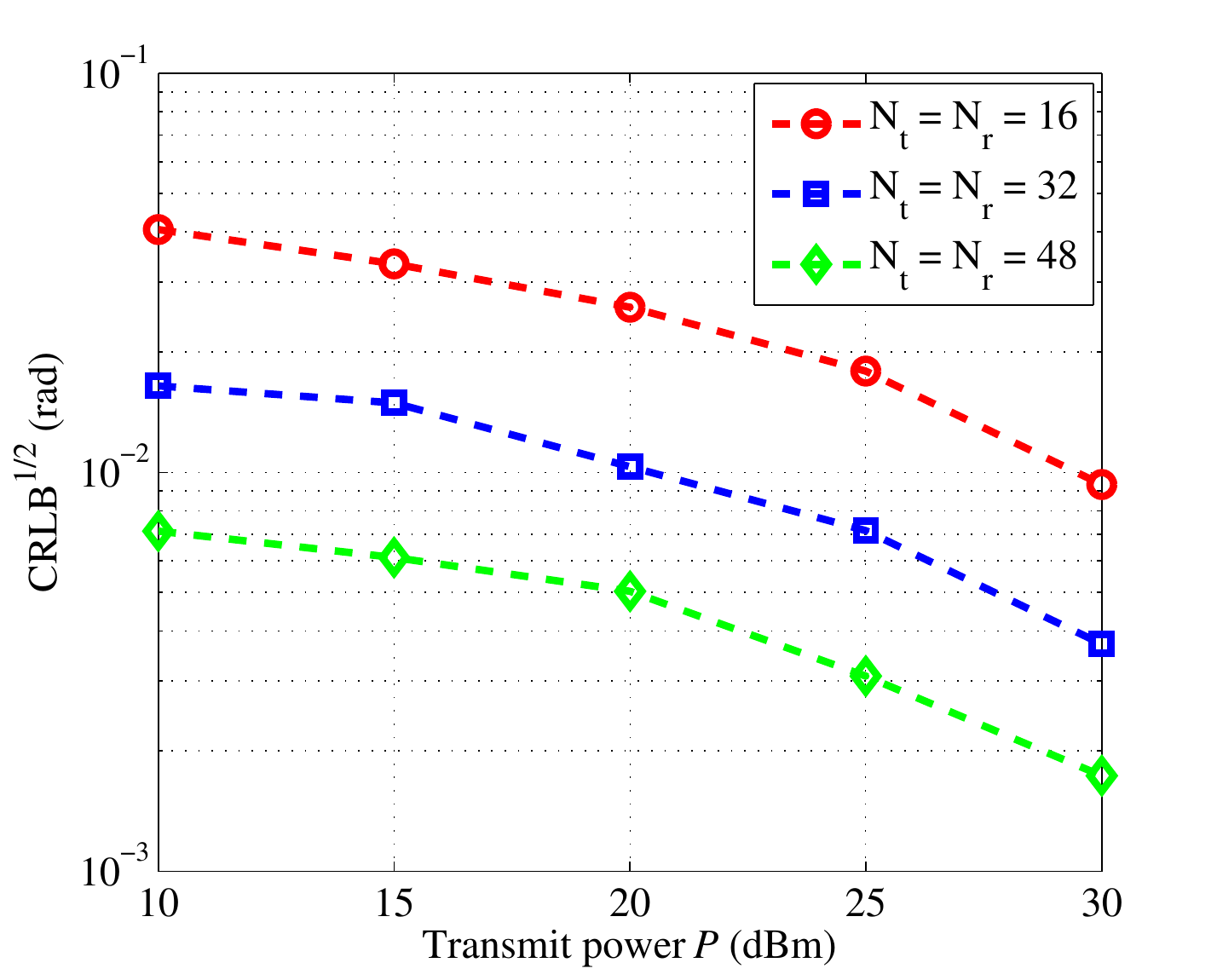}
  \caption{The square root of CRLB achieved by the proposed method in terms of angle estimation with different transmit powers and number of antennas under $\gamma_{\theta} = 0.01~\mathrm{rad}^2$ and $\gamma_d = 0.01~\mathrm{m}^2$.}\label{Fig:Sqrt_CRLB_angle_P}
\end{figure}

\begin{figure}[t]
  \centering
  \includegraphics[width=3in,height=2.6in]{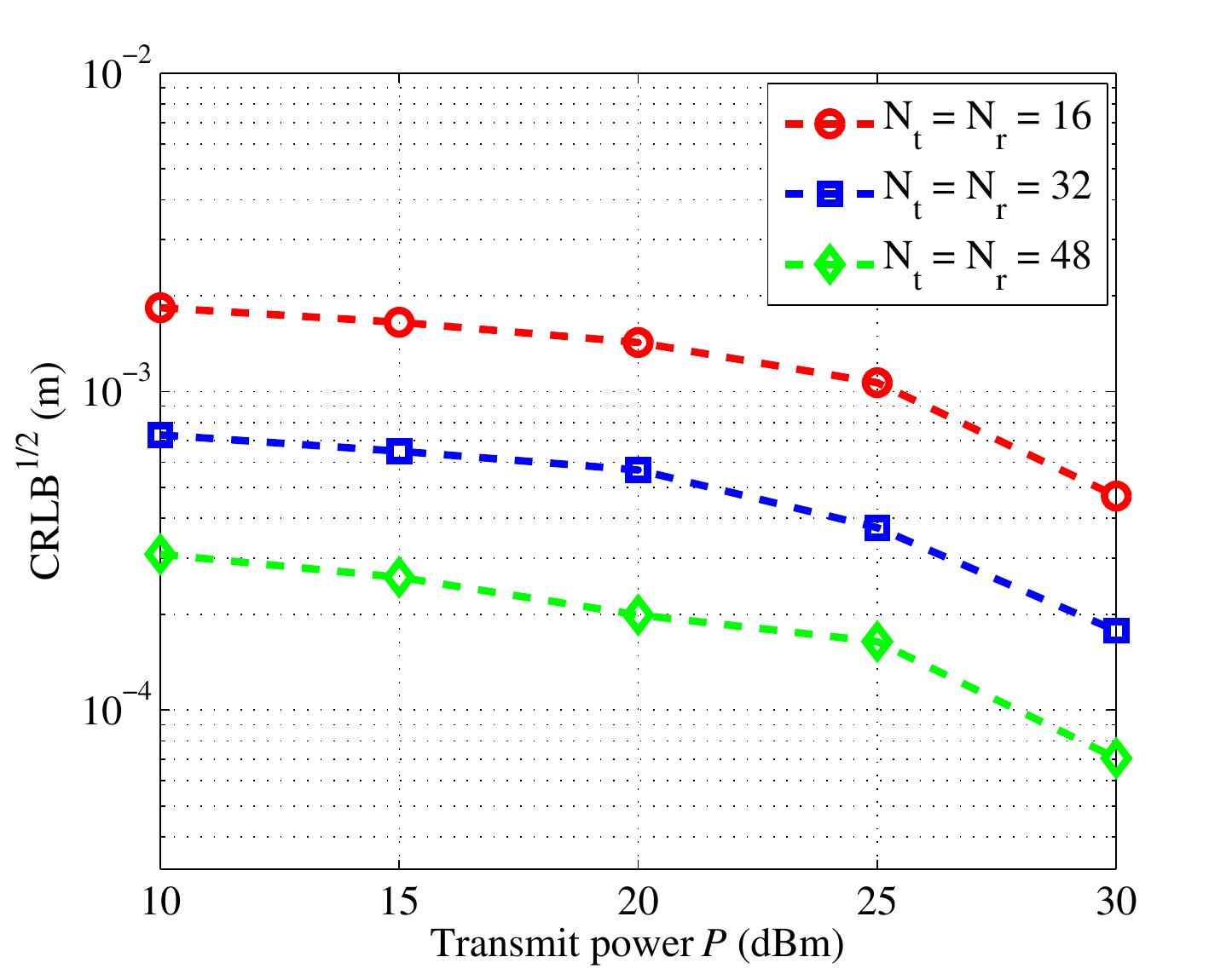}
  \caption{The square root of CRLB achieved by the proposed method in terms of distance estimation with different transmit powers and number of antennas under $\gamma_{\theta} = 0.01~\mathrm{rad}^2$ and $\gamma_d = 0.01~\mathrm{m}^2$.}\label{Fig:Sqrt_CRLB_range_P}
\end{figure}

\subsection{Sensing Performance}
To evaluate the sensing performance of our proposed method, we investigate the achieved tracking performance of vehicles' motion parameters by using the predictive beamforming matrix.
For ease of study, we adopt the square root of CRLB, i.e., the root MSE (RMSE) lower bound of parameter estimation as the metric to characterize the sensing performance in the following simulations.
Denote by $\mathrm{CRLB}^{\frac{1}{2}}$ the square root of CRLB, Fig. \ref{Fig:Sqrt_CRLB_angle_P} illustrates the angle estimation results under different transmit powers and number of antennas, where the maximum tolerable CRLB thresholds in (P1) are set as $\gamma_{\theta} = 0.01~\mathrm{rad}^2$ and $\gamma_d = 0.01~\mathrm{m}^2$, as shown in Table \ref{Tab:Simulation_settings}.
We can observe that all the $\mathrm{CRLB}^{\frac{1}{2}}$ values are around an order of $10^{-3}~\mathrm{rad} \sim 10^{-2}~\mathrm{rad} $, which are much less than the constrained thresholds. Thus, our proposed method can achieve a satisfactory tracking accuracy for V2I networks.
In addition, it is shown that the $\mathrm{CRLB}^{\frac{1}{2}}$ values of all the curves decrease with the increase of $P$.
For example, when $N_t = N_r = 48$ and the transmit power increases from $10~\mathrm{dBm}$ to $15~\mathrm{dBm}$, the $\mathrm{CRLB}^{\frac{1}{2}}$ dramatically decreases from $10^{-2}~\mathrm{rad}$ to $10^{-3}~\mathrm{rad}$.
The reason is that a large transmit power contributes to improving the received SNR at the RSU when receiving the echoes, thus the impact of noise becomes relatively weak and a more accurate angle estimation can be achieved.
Moreover, it can be seen that by increasing the number of antennas at the RSU, the achievable $\mathrm{CRLB}^{\frac{1}{2}}$ is decreasing. It is as expected since large scale antenna array can bring an improved antenna gain to the RSU system.
Based on the results, we can observe that increasing the power budget or number of antennas at the RSU can enhance the received signal strengths at both the RSU and vehicles, thus further improving the sensing and communication performance simultaneously.

Correspondingly, the $\mathrm{CRLB}^{\frac{1}{2}}$ results in terms of distance estimation under different transmit powers and number of antennas are presented in Fig. \ref{Fig:Sqrt_CRLB_range_P}.
It can be observed that when $N_t = N_r = 48$, the proposed method can achieve a $\mathrm{CRLB}^{\frac{1}{2}}$ value of $10^{-4}~\mathrm{m}$, which is sufficient for accurately tracking high-speed vehicles.
Therefore, by exploiting the DL approach to solve the predictive beamforming problem for V2I networks, the proposed method not only performs a satisfactory communication rate but also guarantees the sensing performance.

\begin{figure}[t]
  \centering
  \includegraphics[width=3in,height=2.6in]{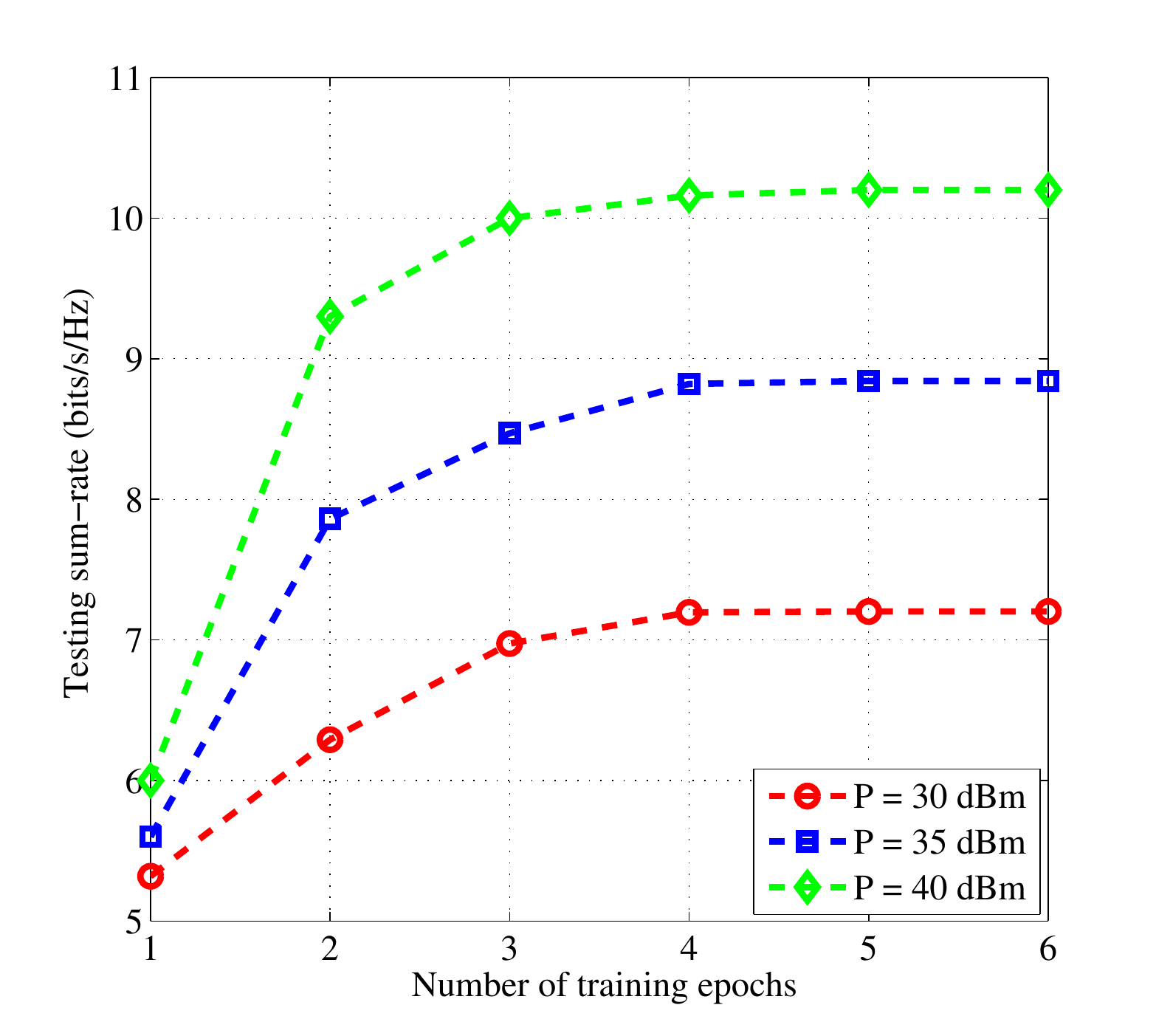}
  \caption{The testing average achievable sum-rate with different epochs under $N_t = N_r = 32$.}\label{Fig:Testing_rate_epoch}
\end{figure}

\begin{figure}[t]
  \centering
  \includegraphics[width=3in,height=2.3in]{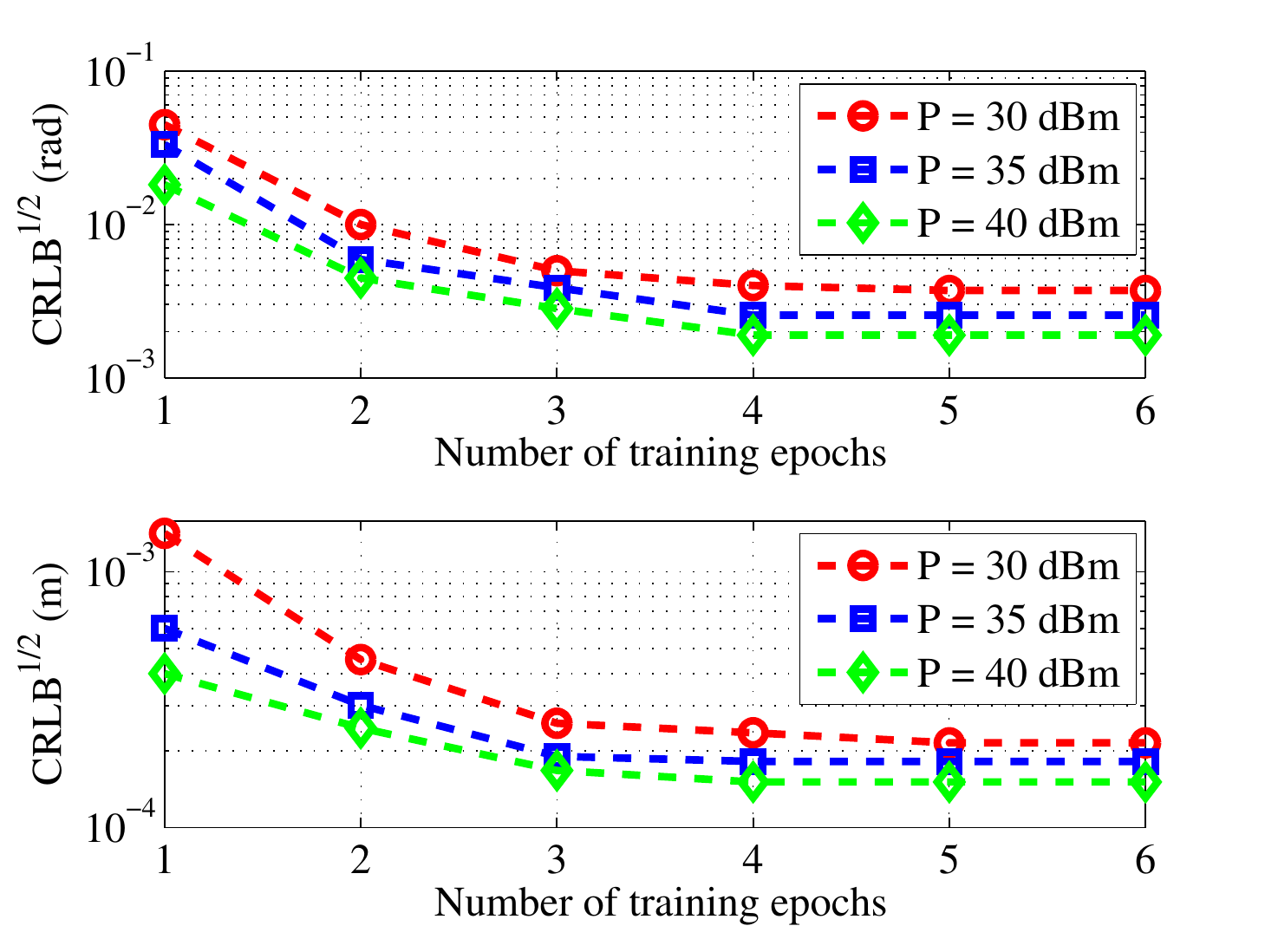}
  \caption{The square root of CRLB with different epochs under $N_t = N_r = 32$: The upper half is the angle estimation performance and the lower half is the distance estimation performance.}\label{Fig:Sqrt_CRLB_epoch}
\end{figure}

\subsection{Training Performance of HCL-Net for ISAC}
To investigate the training performance of the proposed HCL-Net for ISAC, we will evaluate the communication and sensing performance with different number of training epochs in the following.
To demonstrate the effect of neural network training on the communication performance, Fig. \ref{Fig:Testing_rate_epoch} presents the testing sum-rate versus the number of training epochs under $N_t = N_r = 32$.
It can be observed that the testing sum-rate increases with the number of training epochs. Also, for different maximum transmit powers, the proposed HCL-Net can quickly converge to a stable sum-rate within only $6$ training epochs.
The reason is that the adopted convolutional LSTM structure requires a relatively small number of parameters, thus the designed HCL-Net is more efficient for fast network training.
Correspondingly, the testing $\mathrm{CRLB}^{\frac{1}{2}}$ curves of both the angle estimation and the distance estimation under different training epochs are presented in Fig. \ref{Fig:Sqrt_CRLB_epoch} to evaluate the convergence performance of the proposed method in terms of sensing task.
It is shown that the performance of both the angle and distance estimations can converge within $6$ epochs, which demonstrates the practicality of the proposed HCL-Net method.

\begin{figure}[t]
  \centering
  \includegraphics[width=3in,height=2.6in]{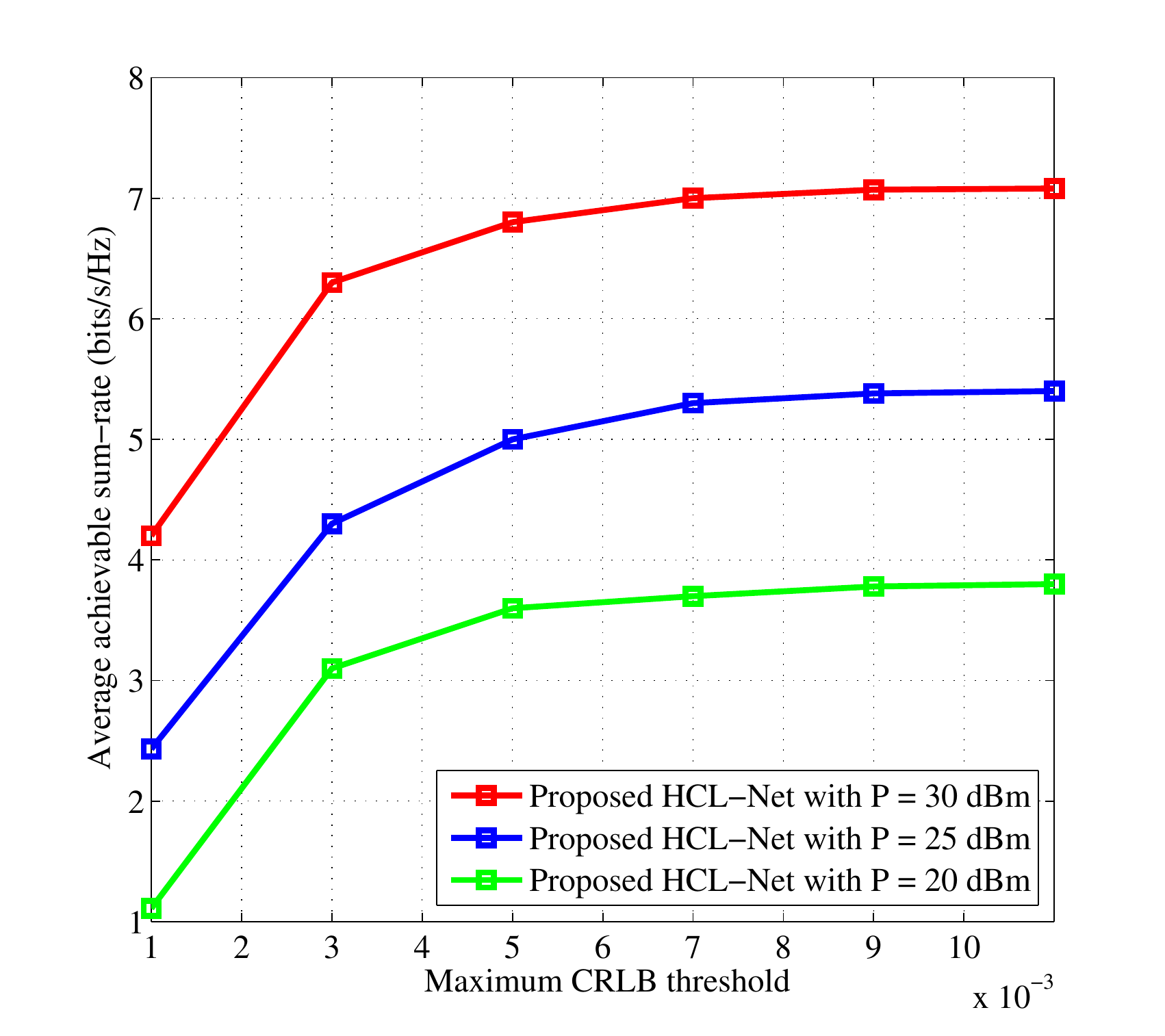}
  \caption{Sensing-communication tradeoff under $N_t = N_r = 32$ and $\gamma_{\theta}=\gamma_d=\gamma$.}\label{Fig:tradeoff}
\end{figure}

\subsection{Sensing-Communication Tradeoff}
To study the interplay between sensing and communication of the proposed method, we plot the maximum tolerable CRLB threshold versus the achievable sum-rate in Fig. \ref{Fig:tradeoff}. We let $\gamma_{\theta}=\gamma_d=\gamma$ and vary the maximum tolerable CRLB threshold $\gamma$ to obtain the corresponding sum-rate.
It can be observed that the sum-rate increases with $\gamma$ and the sum-rate saturates when $\gamma \geq 9\times 10^{-3}$.
This unveils an interesting sensing-communication tradeoff, where if the sensing constraints are loose, the achievable sum-rate can be further improved via beamforming design, otherwise, the sum-rate performance drops.
The results are as expected since the goals of sensing task and communication task are not completely aligned, which makes the desired beamforming matrix to maximize the communication rate differs a lot from that to meet the sensing constraints.
In particular, the sensing-communication tradeoff provides practical guidelines on how to balance the sensing and communication performance in ISAC systems.

\section{Conclusion}
This paper investigated the ISAC in V2I networks and proposed an unsupervised DL-based predictive beamforming scheme to further improve ISAC performance of a V2I system.
First, we formulated a general predictive beamforming problem for ISAC to maximize the communication rate while guaranteeing the sensing accuracy. In particular, we derived the CRLB-based constraints to restrict the tracking performance of vehicles within a required accuracy.
Then a versatile learning-based predictive beamforming framework was developed, which consists of a penalty method-based problem transformation and a DL-based algorithm for solving the problem at hand.
As a realization of the developed predictive framework, a HCL-Net was designed to facilitate the beamforming design of ISAC which is model-free and only requires the historical estimated channels as the network input. In particular, to fully exploit the spatial and temporal features of historical estimated channels, a convolutional LSTM structure was adopted in HCL-Net to further improve the beamforming performance.
Finally, numerical results demonstrated that the proposed HCL-Net can achieve satisfactory performance in terms of both communication and sensing tasks, and the achievable sum-rate of the proposed predictive method approaches the upper bound obtained by a genie-aided scheme exploiting the perfect ICSI.

\bibliographystyle{ieeetr}

\setlength{\baselineskip}{10pt}

\bibliography{ReferenceSCI2}

\end{document}